 \documentclass[preprint2]{aastex}

\usepackage{amsmath}
\usepackage{amssymb}

\newcommand{\rmn}[1]{{\mathrm{#1}}}                  
\newcommand{\bmath}[1]{{\boldsymbol{#1}}}

\shorttitle{Excitation of stellar oscillations by GWs}
\shortauthors{Siegel \& Roth}

\begin{document}


\title{Excitation of stellar oscillations by gravitational waves: hydrodynamic model and numerical results for the Sun}


\author{D. M. Siegel and M. Roth}
\affil{Kiepenheuer-Institut f\"ur Sonnenphysik, Sch\"oneckstr. 6, 79104 Freiburg, Germany}
\email{daniel.siegel@kis.uni-freiburg.de}
\email{markus.roth@kis.uni-freiburg.de}


\begin{abstract}
	Starting from a general relativistic framework a hydrodynamic formalism is derived that yields the mean-square amplitudes and rms surface velocities of normal modes of non-relativistic stars excited by arbitrary gravitational wave (GW) radiation. In particular, stationary GW fields are considered and the resulting formulae are evaluated for two general types of GW radiation: radiation from a particular astrophysical source (e.g., a binary system) and a stochastic background of gravitational waves (SBGW). Expected sources and signal strengths for both types of GW radiation are reviewed and discussed. Numerical results for the Sun show that low-order quadrupolar \textit{g} modes are excited more strongly than \textit{p} modes by orders of magnitude. Maximal rms surface velocities in the case of excitation by astrophysical sources are found to be $v\la 10^{-8}\,\rmn{mm}\,\rmn{s}^{-1}$, assuming GW strain amplitudes of $h\la 10^{-20}$. It is shown that current models for an SBGW produced by cosmic strings, with $\Omega_{\rmn{GW}}\sim 10^{-8}-10^{-5}$ in the frequency range of solar \textit{g} modes, are able to produce maximal solar \textit{g}-mode rms surface velocities of $10^{-5}-10^{-3}\,\rmn{mm}\,\rmn{s}^{-1}$. This result lies close or within the amplitude range of $10^{-3}-1\,\rmn{mm}\,\rmn{s}^{-1}$ expected from excitation by turbulent convection, which is currently considered to be responsible for stellar \textit{g}-mode excitation. It is concluded that studying \textit{g}-mode observations of stars other than the Sun, in which excitation by GWs could be even more effective due to different stellar structures, might provide a new method to either detect GWs or to deduce a significant direct upper limit on an SBGW at intermediate frequencies between the pulsar bound and the bounds from interferometric detectors on Earth.
\end{abstract}





\section{Introduction}

Excitation of solar and solar-like oscillations in non-relativistic stars is generally attributed to turbulent convection. Building on first analytical results by \citet{Goldreich1977}, subsequent theoretical studies (e.g., \citealt{Balmforth1992}; \citealt{Goldreich1994}; \citealt{Samadi2001a,Belkacem2008}) extended and refined the model to describe stochastic excitation of stellar normal modes by turbulent convection and identified two major driving terms in the wave equation. The first source term is related to the Reynolds stress tensor and the second term is due to turbulent entropy fluctuations. However, the latter term was shown to be negligible for excitation of solar \textit{g} modes \citep{Belkacem2009,Appourchaux2010}. Whereas theoretically computed excitation rates for radial solar \textit{p} modes were found to be in good agreement with observations by the GOLF instrument aboard the \textit{Solar and Heliospheric Observatory} (SOHO) spacecraft \citep{Samadi2003a,Samadi2003b,Belkacem2006a,Belkacem2006b}, quantitative estimates for rms surface velocities of solar \textit{g} modes differ from each other by several orders of magnitude (e.g., \citealt{Gough1985,Kumar1996,Belkacem2009}). This is predominantly due to the adopted eddy-time correlation function (see \citealt{Belkacem2009} and the discussion in \citealt{Appourchaux2010}).

In this paper, we build on these earlier works on excitation mechanisms and on recent work by the authors of this paper (\citealt{Siegel2010}, henceforth Paper I) and develop a hydrodynamic model that describes excitation of normal modes of non-relativistic stars by arbitrary, externally generated gravitational waves (GWs). In the past, many authors investigated the possibility of exciting elastic waves in solid astrophysical objects like planets or the Moon by GWs (e.g., \citealt{Forward1961}; \citealt{Weber1967,Dyson1969,Zimmerman1980}) and even excitation of normal modes of fluid stars by GWs was considered (\citealt{Boughn1984,Khosroshahi1997}). However, many of these models are based on highly simplifying assumptions. At the advent of asteroseismology, mainly stimulated by high-precision observations of stellar oscillations from satellite missions like \textit{CoRoT} \citep{Baglin2006} and \textit{Kepler} \citep{Christensen-Dalsgaard2009}, we start to elaborate on the topic again and present here for the first time a thorough hydrodynamic model that describes this excitation mechanism properly. Our model holds for any non-relativistic star and arbitrary external GW fields, it is compatible with current models for other excitation mechanisms (see above) and it allows us to use realistic current solar and stellar models. In fact, it turns out that the formalism presented here is a generalization of the formalism by \citet{Samadi2001a} and \citet{Belkacem2008} to include excitation of normal modes by GWs. Numerical results for the Sun that we present in the second part of the paper suggest that this excitation mechanism together with experimental data from the \textit{Kepler} satellite and theoretical computations in the case of excitation by turbulent convection is likely to either provide a new method of detecting GWs or of deducing a significant upper limit on a stochastic background of gravitational wave radiation (SBGW) at intermediate frequencies in the near future.

The paper is organized as follows. We start from Einstein's field equations of general relativity in Section \ref{sec:hydrodynamic_model} in order to derive an equation of motion for the oscillations of a non-relativistic star under external forcing by GWs. After presenting the general solution to this excitation problem given by an expression for the mean-square amplitudes of normal modes, we focus on stationary external GW fields and derive comparatively simple analytical formulae for mean-square amplitudes and intrinsic rms surface velocities in this case. In Section \ref{sec:GW_radiation}, these formulae are further evaluated for two general types of possible incoming GW radiation: excitation by a particular astrophysical source, such as a binary system, and excitation by an SBGW of astrophysical or cosmological origin. In the second part of this paper, Section \ref{sec:results}, we focus on the Sun as our nearest (non-relativistic) star and present numerical results for the excitation mechanism developed in the preceding sections. In this context, the solar model and damping rates are discussed. We then proceed by reviewing possible astrophysical sources of GW radiation, various models for different candidates of an SBGW as well as expected signal strengths. This finally allows us to numerically compute mean-square amplitudes and intrinsic rms surface velocities for both types of GW radiation; the results of these calculations are presented in Sections \ref{sec:results_as2} and \ref{sec:results_sbgw2}. Although results for solar \textit{p} modes are shown, the focus primarily lies on solar \textit{g} modes, and in particular the low-order ones, since these modes are shown to be more strongly excited than \textit{p} modes and since \textit{p} modes are by far more readily excited by turbulent convection. Conclusions are drawn in Section \ref{sec:conclusion} and two appendices are added to streamline the presentation in Sections \ref{sec:hydrodynamic_model} and \ref{sec:GW_radiation}.

\section{Hydrodynamic model}
\label{sec:hydrodynamic_model}

\subsection{General relativistic framework \label{sec:theoretical_formulation}}

The starting point for the formation of a hydrodynamic theory that describes excitation of normal modes of non-relativistic stars by externally generated GWs is, as in Paper I, the general relativistic conservation equation of energy and momentum\footnote{Throughout this paper, Greek indices take spacetime values 0, 1, 2, 3, whereas Latin indices take spatial values 1, 2, 3 only. Repeated indices are summed over.},
\begin{equation}
	T^{\mu\nu}_{\phantom{\mu\nu};\nu}=0, \label{eq:EMC1}
\end{equation}
or equivalently
\begin{equation}
	\frac{\partial T^{\mu\nu}}{\partial x^\mu}=-\Gamma^\nu_{\mu\sigma}T^{\mu\sigma}-\Gamma^\mu_{\mu\sigma}T^{\sigma\nu}, \label{eq:EMC2}
\end{equation}
which follows directly from Einstein's field equations of general relativity and the Bianchi identities; $T^{\mu\nu}$ denote the components of the energy--momentum tensor of the star under consideration. Here, we assume an ideal fluid, in which case the energy--momentum tensor is given by
\begin{equation}
	T^{\mu\nu}=\left(\rho+\frac{p}{c^2}\right)u^\mu u^\nu +pg^{\mu\nu},
\end{equation}
where the symbols $\rho$, $p$, $c$, and $u^\mu$ are used to denote density, pressure, the speed of light, and the internal four-velocity field, respectively.

For the reason stated in Paper I, we work in Fermi normal coordinates with the origin at the center of mass, $P(t)$, at all times. In this reference frame, the metric $g_{\mu\nu}$ up to second order in the coordinates $x^\mu$ is given by the following expressions (\citealt{ManasseMisner1963}; \citealt{MisnerThorneWheeler1973}):
\begin{eqnarray}
	g_{00}&=&-1-R_{0l0m}x^lx^m \label{eq:FNC1}\\
	g_{0i}&=&-\frac{4}{3}R_{0lim}x^lx^m\\
	g_{ij}&=&\eta_{ij}-\frac{1}{3}R_{iljm}x^lx^m, \label{eq:FNC3}
\end{eqnarray}
where $\eta_{\mu\nu}=\rm{diag}(-1,+1,+1,+1)$ and $R_{\mu\nu\sigma\rho}$ denote the components of the Riemann curvature tensor\footnote{The conventions for the Riemann tensor, the connection, etc., used throughout this paper correspond to the conventions employed by \citet{MisnerThorneWheeler1973} and \citet{Maggiore2008}.} evaluated at $P(t)$, that is along the world line $x^i=0$.

In the case of non-relativistic stars, the linearized theory of general relativity is appropriate by definition. Consequently, the metric can be assumed to be of the form
\begin{equation}
	g_{\mu\nu}=\eta_{\mu\nu}+h_{\mu\nu},
\end{equation}
with $h_{\mu\nu}$ being very small, $\vert h_{\mu\nu}\vert\ll 1$, and stars are taken to move in flat spacetime, i.e., deviations from flat space are solely due to GWs, $h_{\mu\nu}=h_{\mu\nu}^{\rm{GW}}$, and effects of curvature due to the mass of the star are neglected. To linear order in $h_{\mu\nu}$, the components of the Riemann tensor in linearized theory are given by
\begin{multline}
	R_{\mu\nu\rho\sigma}=\frac{1}{2}(\partial_{\nu}\partial_{\rho}h_{\mu\sigma}+\partial_{\mu}\partial_{\sigma}h_{\nu\rho}\\
	-\partial_{\mu}\partial_{\rho}h_{\nu\sigma}-\partial_{\nu}\partial_{\sigma}h_{\mu\rho}). \label{eq:Riemann}
\end{multline}
It is important to stress that the Riemann tensor in linearized theory is invariant, rather than just covariant, and thus it can be evaluated in any coordinate frame. For convenience, we choose $h_{\mu\nu}$ to be in transverse-traceless (TT) gauge defined by
\begin{equation}
	h_{0\mu}=\partial^jh_{ij}=h^i_{\phantom{i}i}=0.  \label{eq:tt_gauge}
\end{equation}

As in Paper I, there are some well-motivated approximations to be made. First, we adopt the long-wavelength approximation, i.e., it is assumed that the typical length scale $\lambda/2\pi$, over which the incident gravitational radiation with wavelength $\lambda$ changes substantially, is much larger than the diameter $d$ of the star, $d/(\lambda/2\pi)\ll 1$. This results in $h_{\mu\nu}$ having no spatial dependence over the volume of the star. Second, we assume that the internal motions of the star are purely Newtonian, i.e., non-relativistic. Consequently, only terms including $T^{00}$ need be retained on the right-hand side of Equation (\ref{eq:EMC2}), with $T^{00}$ given by $T^{00}=\rho c^2$.

\subsection{Equation of motion \label{sec:EOM}}

In order to derive an equation of motion from Equation (\ref{eq:EMC1}) or (\ref{eq:EMC2}) that describes the dynamics of the internal velocity field of the star under external forcing by GWs, the components $T^{\mu\nu}$ of the energy--momentum tensor and the Christoffel symbols have to be evaluated in Fermi normal coordinates. This is done by expressing these quantities in terms of the metric components given by Equations (\ref{eq:FNC1})--(\ref{eq:FNC3}). Due to their invariance property, the components of the Riemann tensor can then be computed with the help of Equations (\ref{eq:Riemann}) and (\ref{eq:tt_gauge}). Finally, adopting the aforementioned approximations and only retaining terms linear in the coordinates $x^\mu$ and linear in the metric perturbations $h_{\mu\nu}$ yields the usual equation of continuity as the zeroth component of Equation (\ref{eq:EMC2}),
\begin{equation}
	\frac{\partial\rho}{\partial t}+\bmath{\nabla\cdot}(\rho \bmath{v})=0, \label{eq:continuity}
\end{equation}
where $\bmath{v}$ denotes the internal velocity field of the star, and a modified Euler equation,
\begin{equation}
	\frac{\partial\rho\bmath{v}}{\partial t}+\bmath{\nabla\cdot}(\rho\bmath{v}\otimes\bmath{v})=-\nabla p+\bmath{f}_{\rm{GW}}, \label{eq:Euler1}
\end{equation}
which originates from the spatial components of Equation (\ref{eq:EMC2}). The latter equation has an additional term compared to the usual Euler equation of hydrodynamics, $\bmath{f}_{\rm{GW}}(\bmath{x},t)$, with components given by
\begin{equation}
	f_{\rm{GW}}^i=\frac{1}{2}\rho \ddot{h}^i_{\phantom{i}j} x^j\equiv\rho\bar{f}^i_{\rmn{GW}}.
\end{equation}
This term represents an additional external driving force per unit volume exerted by GWs. Detailed derivations of Equations (\ref{eq:continuity}) and (\ref{eq:Euler1}) are given in Appendix \ref{sec:app_1}.

Effects of the gravitational field of the star itself can now be included by adding a Newtonian potential term, $\rho\bmath{g}$. Thus, we finally arrive at
\begin{equation}
	\frac{\partial\rho\bmath{v}}{\partial t}+\bmath{\nabla\cdot}(\rho\bmath{v}\otimes\bmath{v})=-\nabla p+\rho\bmath{g}+\bmath{f}_{\rm{GW}}.\label{eq:Euler2}
\end{equation}
Here, $\bmath{g}$ must satisfy Poisson's equation,
\begin{equation}
	\bmath{\nabla\cdot g}=-4\pi G \rho,
\end{equation}
where $G$ denotes the gravitational constant.

Equation (\ref{eq:Euler2}) describes all internal motions of a non-relativistic star under external forcing by GWs. In the following, we closely follow the procedure presented by \citet{Samadi2001a} in order to derive an equation of motion solely for the oscillations of the star that are excited in this way. As a by-product of applying this procedure, we obtain a generalized and unified theoretical description of the mode excitation processes due to turbulent convection and GWs, combined in a single consistent forced wave equation. However, in contrast to \citet{Samadi2001a} and previous work by \citet{Goldreich1977} and \citet{Goldreich1994}, which the former is based on, we do not adopt the Cowling approximation, that is we do not neglect perturbations in the Newtonian gravitational potential. Instead, we show that terms related to these perturbations can be retained as far as oscillations are concerned.

All physical quantities are decomposed into a sum of the equilibrium quantity, $q_0$, and an Eulerian fluctuation, $q_1$, $q=q_0+q_1$. We retain terms up to quadratic order in the fluctuations $p_1$, $\rho_1$, $s_1$, $\bmath{g}_1$, and $\bmath{v}$. However, since in calculations of stellar normal modes only terms up to linear order in the perturbations are consistently retained, we neglect nonlinear terms that neither appear as damping nor as driving terms in the resulting equation of motion. Proceeding in this way, the perturbed equations that we consider are written as
\begin{equation}
	\frac{\partial \rho_1}{\partial t}+\bmath{\nabla\cdot}(\rho\bmath{v})=0, \label{eq:continuity_per}%\\
\end{equation}
\begin{multline}	
	\frac{\partial \rho \bmath{v}}{\partial t}+\nabla p_1-\rho_1\bmath{g}_0+\rho_0\bmath{g}_1\\
	=-\bmath{\nabla\cdot}(\rho_0 \bmath{v}\otimes\bmath{v})+\rho \bar{\bmath{f}}_{\rmn{GW}},\label{eq:Euler_per}%\\
\end{multline}
\begin{equation}
	\bmath{\nabla\cdot} \bmath{g}_1=-4\pi G \rho_1. \label{eq:Poisson_per}
\end{equation}
%\end{eqnarray}  
Additionally, from the assumption of adiabatic oscillations one has
\begin{equation}
	\frac{ds_t}{dt}=\frac{\partial s_1}{\partial t}+\bmath{v\cdot\nabla}(s_0+s_1), \label{eq:entropy}
\end{equation}
where $s_t$ denotes the Lagrangian entropy fluctuation due to turbulent convection \citep{Samadi2001a}.

Equations (\ref{eq:continuity_per}), (\ref{eq:Poisson_per}), and (\ref{eq:entropy}) are now employed to derive the desired equation of motion for the oscillation modes from the temporal derivative of Equation (\ref{eq:Euler_per}). The velocity field $\bmath{v}$ is separated into a component $\bmath{v}_{\rmn{osc}}$ due to oscillations and a component $\bmath{u}$ due to all other motions, the most prominent of which is turbulent convection,
\begin{equation}
	\bmath{v}=\bmath{v}_{\rmn{osc}}+\bmath{u}. \label{eq:vel_sep}
\end{equation}
Equation (\ref{eq:vel_sep}) implies that the turbulent medium is decoupled from the oscillation modes. This approximation is essential in treating the excitation problem analytically, since it is ultimately responsible for the resulting equation of motion being a linear, inhomogeneous and thus analytically solvable differential equation. Furthermore, we assume incompressible turbulence, $\bmath{\nabla\cdot u}=0$. The result is the inhomogeneous wave equation for the oscillation modes,
\begin{multline}
	\rho_0\left(\frac{\partial^2}{\partial t^2}-\mathcal{L}\right)\bmath{v}_{\rm{osc}}+\mathcal{D}(\bmath{v}_{\rm{osc}})\\
	=\frac{\partial}{\partial t}(\bmath{f}_{\rm{Rey}}+\bmath{f}_{\rm{entr}}+\bmath{f}_{\rm{GW}}), \label{eq:EOM1}
\end{multline}
where the wave operator $\mathcal{L}$ is a linear differential operator acting on arbitrary differentiable vector fields $\bmath{X}$ according to
\begin{multline}
	\mathcal{L}(\bmath{X})\equiv \frac{1}{\rho_0}\bigg\{\nabla\left[c_s^2\bmath{\nabla\cdot}(\rho_0\bmath{X})+\alpha_s\bmath{X\cdot\nabla} s_0\right]\\
	-\bmath{g}_0\bmath{\nabla\cdot}(\rho_0\bmath{X})\bigg\}-G\nabla_{\bmath{x}}\left[\int_V d^3r\,\frac{\bmath{\nabla\cdot}(\rho_0\bmath{X})}{|\bmath{x}-\bmath{r}|}\right].
\end{multline}
Here, $V$ denotes the volume of the star under consideration. The linear damping operator $\mathcal{D}$ is identical to the one obtained by \citet{Samadi2001a} and it is given by Equation (13) therein. On the right-hand side of Equation (\ref{eq:EOM1}) there are three driving terms, the first two of which are, respectively, the Reynold source term, which is due to turbulent Reynolds stress, and the entropy source term, which is due to turbulent entropy fluctuations; definitions of these terms are provided by Equations (11) and (12) of \citet{Samadi2001a}. For the present analysis, however, we are solely interested in the third driving term,
\begin{equation}
	\bmath{f}_{\rmn{GW}}=\rho_0\bar{\bmath{f}}_{\rmn{GW}},
\end{equation}
which represents excitation of normal modes by external GWs; here we neglected the contribution $\rho_1\bar{\bmath{f}}_{\rmn{GW}}$, since it is comparatively small. Consequently, we simply ignore the first two driving terms. This, however, is only possible because of the source terms being independent of the terms on the left-hand side of the equation, thanks to the separation of the velocity field into an oscillatory and a turbulent component. Hence, the final equation of motion to be solved is written as
\begin{equation}
	\rho_0\left(\frac{\partial^2}{\partial t^2}-\mathcal{L}\right)\bmath{v}_{\rm{osc}}+\mathcal{D}(\bmath{v}_{\rm{osc}})=\frac{\partial\bmath{f}_{\rm{GW}}}{\partial t}. \label{eq:EOM2}
\end{equation}
We note the striking formal similarity of Equation (\ref{eq:EOM2}) with the equation of motion derived in Paper I in the case of adopting the isotropic elastic sphere model of a star, which corresponds to considering the Sun as a spherical resonant mass GW detector.

\subsection{Square amplitudes and surface velocities \label{sec:sq_ampl,surf_vel}}

This subsection is dedicated to solving the equation of motion (\ref{eq:EOM2}) and to thereby gaining analytical formulae for the square amplitudes and surface velocities of global stellar normal modes excited by a not yet to be specified external GW field, which is represented by the external driving force $\bmath{f}_{\rm{GW}}$. Henceforth, we will drop the subscript ``osc'' and solely use $\bmath{v}$ to denote the velocity field due to oscillations.

We start by discussing the corresponding homogeneous equation without driving and damping terms,
\begin{equation}
	\left(\frac{\partial^2}{\partial t^2}-\mathcal{L}\right)\bmath{v}_{\rm{hom}}=0, \label{eq:hom_EOM}
\end{equation}
in which case the velocity field can be assumed to have harmonic time dependence,
\begin{equation}
	\bmath{v}_{\rmn{hom}}(\bmath{x},t)=-i\omega\bmath{\xi}(\bmath{x})e^{-i\omega t},
\end{equation}
where $\omega$ is the oscillation frequency and $\bmath{\xi}(\bmath{x})$ is a function that absorbs the spatial dependence of $\bmath{v}_{\rmn{hom}}$. Inserting this ansatz into Equation (\ref{eq:hom_EOM}) yields the usual eigenvalue problem,
\begin{equation}
	\mathcal{L}\bmath{\xi}(\bmath{x})=-\omega^2\bmath{\xi}(\bmath{x}),
\end{equation}
which, subject to appropriate boundary conditions, can be numerically solved by stellar pulsation codes (e.g., \citealt{ChristensenDalsgaard2008}). The eigenfunctions $\{\bmath{\xi}_N(\bmath{x})\}$ of $\mathcal{L}$ can be shown to be orthogonal with respect to the inner product defined by
\begin{equation}
	\langle \bmath{X},\bmath{Y}\rangle\equiv\int_V d^3x\,\rho_0\bmath{X}^*(\bmath{x})\bmath{\cdot}\bmath{Y}(\bmath{x}),
\end{equation}
see, e.g., \citet{Unno}. Here $*$ denotes complex conjugation and $V$ denotes the volume of the star under consideration. Furthermore, the differential operator $\mathcal{L}$ is Hermitian \citep{Unno} and it is generally assumed that its eigenfunctions form a complete set (cf., e.g., \citealt{Aizenman1977}). As shown in Paper I, toroidal eigenmodes do not couple to external GWs and additionally these modes are of no relevance to asteroseismology, since they do not perturb the hydrostatic equilibrium and therefore have eigenfrequencies $\omega=0$ \citep{Aizenman1977}. This allows us to solely focus on the usual spheroidal eigenmodes,
\begin{equation}
	\bmath{\xi}_{nlm}(r,\Theta,\phi)=\left[\xi_{r,nl}(r)\bmath{e}_r+\xi_{h,nl}(r)r\nabla\right]Y_{lm}(\Theta,\phi), \label{eq:ef}
\end{equation} 
which we normalize according to
\begin{equation}
	\int_V d^3x\, \rho_0 \bmath{\xi}_N^*\bmath{\cdot}\bmath{\xi}_{N'}=I\delta_{NN'}, \label{eq:normalization}
\end{equation}  
where $N\equiv\{nlm\}$ is an abridged index and $I$ is a constant with dimension of a mass. We arbitrarily normalize the eigenfunctions to unit mass in cgs units, that is, $I\equiv 1\,\rmn{g}$.

The previous discussion shows that the velocity field in Equation (\ref{eq:EOM2}) can be expanded as
\begin{equation}
	\bmath{v}(\bmath{x},t)=\sum_N (-i\omega_N) A_N(t)\bmath{\xi}_N(\bmath{x})e^{-i\omega_N t}, \label{eq:vel_expan}
\end{equation}
where the time-dependent complex expansion coefficients $A_N(t)$ are called amplitudes and where only eigenfunctions of the form (\ref{eq:ef}) need be considered. Furthermore, we assume weak damping, i.e., $\eta_N\ll\omega_N$, where $\eta_N$ is the damping rate. Thus, we have $|\partial A_N/\partial t|\ll\omega_N |A_N|$. In order to obtain analytical expressions for the amplitudes $A_N$, we proceed as in Paper I and solve Equation (\ref{eq:EOM2}) for each velocity eigenmode,
\begin{equation}
	\bmath{v}_N(\bmath{x},t)\equiv-i\omega_N A_N(t)\bmath{\xi}_N(\bmath{x})e^{-i\omega_N t}, \label{eq:vel_ef}
\end{equation}
separately. Inserting the ansatz (\ref{eq:vel_ef}) into Equation (\ref{eq:EOM2}) and adopting the weak-damping assumption yields the expression
\begin{equation}
	\rho_0\bmath{\xi}_N\frac{dA_N}{dt}+\frac{i}{2\omega_N}\mathcal{D}_N'(\bmath{\xi}_N)A_N=\frac{-e^{i\omega_N t}}{2\omega_N^2}\frac{\partial\bmath{f}_{\rmn{GW}}}{\partial t}. \label{eq:Sol}
\end{equation}
The differential damping operator $\mathcal{D}$ may in general contain time derivatives (cf. Equation (13) of \citealt{Samadi2001a}) and thus the resulting operator $\mathcal{D}'_N$ after applying the weak-damping assumption and factoring out $A_Ne^{-i\omega_N t}$ depends on the particular eigenmode $N$ through the eigenfrequency $\omega_N$. Multiplying Equation (\ref{eq:Sol}) by $\bmath{\xi}_N^*$, integrating over the volume $V$ of the star, employing the normalization condition (\ref{eq:normalization}), and rearranging yields
\begin{equation}
	\frac{dA_N}{dt}+\sigma_N A_N =\frac{-e^{i\omega_N t}}{2\omega_N^2I}\int_V d^3x\,\bmath{\xi}_N^*\bmath{\cdot}\frac{\partial\bmath{f}_\rmn{GW}}{\partial t}, \label{eq:EOM_A_N}
\end{equation}
where we defined
\begin{equation}
	\sigma_N\equiv \frac{i}{2\omega_N I}\int_V d^3x\,\bmath{\xi}_N^*\bmath{\cdot}\mathcal{D}_N'(\bmath{\xi}_N)\equiv\eta_{D,N}+i\delta\omega_N.
\end{equation}
It is worthwhile to note that the damping operator $\mathcal{D}$ only includes dynamical sources of damping, since the mode energy equation is not considered. Therefore, the real part $\eta_{D,N}$ of $\sigma_N$ in Equation (\ref{eq:EOM_A_N}) will contribute to dynamical linear mode damping. This is why we will henceforth replace $\eta_{D,N}$ by a total damping rate $\eta_N$ to include possible additional damping mechanisms not yet considered in Equation (\ref{eq:EOM2}). The damping operator gives also rise to a dynamical shift of the oscillation frequency, which is manifested in the imaginary part $\delta\omega_N$ of $\sigma_N$. These frequency shifts are generally expected to be smaller or on the order of the frequency shifts that would result from a non-adiabatic calculation (cf. \citealt{Samadi2001a}) and hence they need not be considered here. Consequently, we set $\sigma_N=\eta_N$. 

Equation (\ref{eq:EOM_A_N}) is of the same form as the corresponding Equation (23) in Paper I. Thus we can apply the same procedure as presented in Section 5 and Appendix A of Paper I to solve this differential equation, thereby obtaining the following general result for the mean-square amplitude:
\begin{eqnarray}
	\big\langle|A_N(t)|^2\big\rangle\mskip-10mu&=&\mskip-10mu\frac{R^8\chi_n^2}{16\omega_N^2 I^2}\int_{-\infty}^{t}dt_0\,
	e^{2\eta_N(t_0-t)}\nonumber\\
	&&\mskip-10mu\times\int_{-2(t-t_0)}^{2(t-t_0)}d\tau\,\langle K_m^{t_0}(\tau)\rangle e^{-i\omega_N \tau},\nonumber\\ 
	\label{eq:sq_ampl_gen}
\end{eqnarray}
where $R$ denotes the radius of the star and $\langle\rangle$ denotes an ensemble average, i.e., an expectation value. In solving Equation (\ref{eq:EOM_A_N}) explicit use of the weak-damping assumption and the long-wavelength approximation is made. Furthermore, it is assumed that the driving term $\bmath{f}_{\rmn{GW}}$ is ``turned on'' at $t=-\infty$ and that one is interested in the case of large $t$, i.e., in the late-time behavior of the system, rather than in its transient behavior. It is important to note that, as shown in Appendix A of Paper I, $A_N$ is only non-vanishing for quadrupolar ($l=2$) eigenmodes, that is, in Equation (\ref{eq:sq_ampl_gen}) we have $N=\{n2m\}$ and $\big\langle|A_N(t)|^2\big\rangle\equiv0$ otherwise. Furthermore, in Equation (\ref{eq:sq_ampl_gen}) some of the definitions of Paper I have been transferred and adapted to the present hydrodynamic formalism: the quantity $\chi_n$ is defined by the following integral:
\begin{equation}
	\chi_n\equiv\int_{0}^{1} dz\,\rho_0(z)z^3\left[\xi_{r,n2}(z)+3\xi_{h,n2}(z)\right], \label{eq:chi}
\end{equation}
and moreover, we defined the correlation function
\begin{equation}
	K_m^{t_0}(\tau)\equiv\ddot{h}_m\left(t_0-\frac{\tau}{2}\right)\ddot{h}_m^*\left(t_0+\frac{\tau}{2}\right), \label{eq:corrfunc}
\end{equation}
where $h_m$ denote the spherical components of the GW tensor (a definition is given by Equation (\ref{eq:expan_sph_harm}) in Appendix \ref{sec:app_2}).

If the surrounding GW field is stationary, that is, all statistical quantities of the GW field are time independent and solely depend upon time differences, the amplitudes become time independent, thanks to $K_m^{t_0}(\tau)$ being independent of $t_0$ in this case. As shown in Paper I, the mean-square amplitudes are then given by the following comparatively simple expression:
\begin{equation}
	\big\langle|A_N|^2\big\rangle=\frac{R^8\chi_n^2}{32\eta_{N}\omega_{N}^2 I^2}P_m(\omega_{N}). \label{eq:sq_ampl}
\end{equation}
The second factor in this formula, $P_m(\omega)$, is the power spectral density of the spherical accelerations $\ddot{h}_m$ of the GW tensor,
\begin{equation}
	P_m(\omega)\equiv\int_{-\infty}^{\infty} d\tau\,\langle K_m(\tau)\rangle\,e^{-i\omega\tau}, \label{eq:psd}
\end{equation}
and thus this factor solely depends on the surrounding GW field. It is worth noting that, by contrast, the first factor on the right-hand side of Equation (\ref{eq:sq_ampl}) only depends on the stellar model and quantities derived therefrom. Consequently, the latter quantity describes the star's general ability to resonate in a particular eigenmode $N$ when the star is excited by GWs, and this is why we christen it \textit{resonance factor}.

As in Paper I, we define the corresponding intrinsic mean-square surface velocity of a particular eigenmode by
\begin{eqnarray}
	v_{N,s}^2&\equiv&\bigg\langle\frac{1}{2}\int d\Omega\,\bmath{v}_N(\bmath{x},t)\bmath{\cdot} \bmath{v}_N^*(\bmath{x},t)\bigg\rangle(R)\nonumber\\
	&=&\frac{1}{2}\omega_N^2\big\langle|A_N(t)|^2\big\rangle\Psi_N^2(R), \label{eq:def_sq_surf_vel}
\end{eqnarray}
where 
\begin{equation}
	\Psi_N^2(r)\equiv \xi_{r,n2}^2(r)+l(l+1)\xi_{h,n2}^2(r), \label{eq:psi}
\end{equation}
with $l(l+1)=6$, $\langle\rangle$ denotes time average and $\Omega$ denotes the solid angle. We note that the definition (\ref{eq:def_sq_surf_vel}) does not take visibility factors into account, such as geometrical and limb darkening effects, and it is equivalent to the one employed by \citet{Belkacem2009}. Inserting Equation (\ref{eq:sq_ampl}) into Equation (\ref{eq:def_sq_surf_vel}) yields the following expression for the rms surface velocities:
\begin{equation}
	v_{N,s}=\left[\frac{R^8\chi_n^2}{64\eta_N I^2}P_m(\omega_N)\Psi_N^2(R)\right]^\frac{1}{2}. \label{eq:rms_surf_vel}
\end{equation}
Finally, it is interesting to note that the square amplitudes as well as the mean-square surface velocities show a characteristic $R^8$-dependence on the radius of the star. However, this sensitive dependence on the radius is reduced to an effective $R^2$-dependence due to the adopted normalization scheme (cf. Equation (\ref{eq:normalization})).

\section{Sources of gravitational radiation}
\label{sec:GW_radiation}
In this section, the power spectral density $P_m(\omega)$ defined in Equation (\ref{eq:psd}) is calculated for two general types of GW radiation: radiation from astrophysical sources that can be individually ``resolved'' and that emit approximately monochromatic GW radiation, and an SBGW of astrophysical or cosmological origin. We start with discussing the former case.

\subsection{Excitation by distinct astrophysical sources \label{sec:GW_radiation1}}

In the following, it is assumed that the particular source that emits approximately monochromatic radiation is sufficiently far away from the star under consideration, such that the incoming gravitational radiation can be appropriately described by a plane wave field as defined below. We assume that the incoming plane wave of frequency $\omega$ has its wave vector $\bmath{k}$ aligned with the \textit{z}-axis, $\bmath{k}=|\bmath{k}|\bmath{e}_z$. In this case, the GW tensor in TT gauge (cf. Equation (\ref{eq:tt_gauge})) takes the form
\begin{eqnarray}
	h_{ij}(t)&=&\left( \begin{array}{ccc}
	A_+ & A_\times & 0 \\
	A_\times & -A_+ & 0 \\
	0 & 0 & 0
	\end{array} \right)_{ij} e^{-i\omega t} \label{eq:pw}\\
	&\equiv& h_+ e_{ij}^+(\bmath{e}_z)+h_\times e_{ij}^\times (\bmath{e_z}),\nonumber
\end{eqnarray}
where the long-wavelength approximation (cf. Section \ref{sec:theoretical_formulation}) has already been adopted. The polarization tensors $e_{ij}^+(\bmath{e}_z)$ and $e_{ij}^\times (\bmath{e}_z)$ are defined in Appendix \ref{sec:app_2}. In order to make this formalism directly compatible with waveforms of gravitational radiation from binary systems, we assume that $A_+$ and $A_\times$ are in general complex numbers. This is an important detail, which we make use of in Section \ref{sec:results_as}. Finally, we adopt the usual convention for the GW tensor that the real part is taken at the end of the calculation.

The spherical components $\ddot{h}_m$ of $\ddot{h}_{ij}$, defined by
\begin{equation}
		\ddot{h}_{ij}=\sum_{m=-2}^{2}\ddot{h}_m\mathcal{Y}_{ij}^{2m}
\end{equation}
(cf. Appendix \ref{sec:app_2}), are determined with the help of Equation (\ref{eq:inv_expan_sph_harm}), where $h_{ij}$ is given by Equation (\ref{eq:pw}). One immediately finds: $\ddot{h}_0=0$, $\ddot{h}_{\pm1}=0$, and 
\begin{equation}
	\ddot{h}_{\pm 2}=\sqrt{\frac{8\pi}{15}}(-\omega^2)(A_+ \mp iA_\times)e^{-i\omega t}. \label{eq:hpm2}
\end{equation}
Consequently, gravitational radiation propagating along the \textit{z}-direction solely couples to internal normal modes of the star having an azimuthal order of $m=\pm 2$.

Knowledge of $\ddot{h}_{\pm 2}$ now allows for the computation of the correlation function $K^{t_0}_{\pm2}(\tau)$. Inserting Equation (\ref{eq:hpm2}) into Equation (\ref{eq:corrfunc}) we find after some elementary transformations
\begin{multline}
	K_{\pm{2}}^{t_0}(\tau)=\frac{8\pi}{15}\omega^4\big[|A_+|^2+|A_\times|^2\\
	\pm2\rmn{Re}(iA_+A_\times^*)\big]e^{i\omega \tau}. \label{eq:K}
\end{multline}
We note the high sensitivity of $|K_{\pm{2}}^{t_0}(\tau)|$ to the frequency $\omega$ given by a $\omega^4$ power law. Finally, by Fourier-transforming $K_{\pm{2}}^{t_0}(\tau)$ in accordance with Equation (\ref{eq:psd}), one obtains the power spectral density of $\ddot{h}_m$,
\begin{multline}
	P_{\pm2}(\omega_N)=\frac{16\pi^2}{15}\omega^4\big[|A_+|^2+|A_\times|^2\\
	\pm2\rmn{Re}(iA_+A_\times^*)\big]\delta(\omega-\omega_N). \label{eq:Pm_pw}
\end{multline}

As can be read off Equation (\ref{eq:K}), the GW field is indeed stationary, i.e., the correlation function is independent of $t_0$, and thus Equations (\ref{eq:sq_ampl}) and (\ref{eq:rms_surf_vel}) apply. Resulting square amplitudes and intrinsic rms surface velocities of excited normal modes ($N=\{n,2,\pm2\}$) are, respectively, given by
\begin{multline}
|A_N|^2=\frac{\pi^2}{30}\frac{R^8\chi_n^2\omega_N^2}{\eta_N I^2}\big[|A_+|^2+|A_\times|^2\\
\pm2\rmn{Re}(iA_+A_\times^*)\big] \label{eq:sq_ampl_pw}
\end{multline}
and
\begin{multline}
v_{N,s}^2=\frac{\pi^2}{60}\frac{R^8\chi_n^2\omega_N^4}{\eta_N I^2}\big[|A_+|^2+|A_\times|^2\\
\pm2\rmn{Re}(iA_+A_\times^*)\big]\Psi_N^2(R), \label{eq:surf_vel_pw}
\end{multline}
where for each mode $N$ it was assumed that there is a GW source at the same frequency, $\omega=\omega_N$. Numerical results for Equations (\ref{eq:sq_ampl_pw}) and (\ref{eq:surf_vel_pw}) in the case of the Sun and in view of present amplitude estimates for known and expected individual astrophysical GW sources are presented and discussed in Section \ref{sec:results_as}.

\subsection{Excitation by a stochastic background of gravitational waves \label{sec:GW_radiation2}}

In this subsection, an arbitrary SBGW is considered and no assumption about its origin is made at this point, that is, we do not yet focus on a specific background, such as a background due to cosmic strings or due to astrophysical sources. This step is postponed to Section \ref{sec:results_sbgw1}.

It is convenient to characterize a stochastic background by the dimensionless function
\begin{equation}
	\Omega_{\rmn{GW}}(\nu)\equiv\frac{1}{\rho_c}\frac{d\rho_{\rmn{GW}}}{d\ln\nu}, \label{eq:Om}
\end{equation}
which measures the energy density of GWs per unit logarithmic interval of frequency in units of the present critical energy density, $\rho_c$, that is needed to close the universe,
\begin{equation}
	\rho_\rmn{c}=\frac{3c^2H_0^2}{8\pi G}. \label{eq:rho_c}
\end{equation}
Here $H_0$ denotes the present Hubble expansion rate, $c$ the speed of light, and $G$ the gravitational constant. In the following, we assume that the background is
\begin{enumerate}
	\item[(i)] isotropic,
	\item[(ii)] unpolarized,
	\item[(iii)] stationary, and
	\item[(iv)] Gaussian.
\end{enumerate}
Each of these assumptions is well motivated in the case of an SBGW \citep{Allen1999,Maggiore2008}. In particular, the stationarity property of the background is important, since it allows us to apply the formulae for the mean-square amplitudes and rms surface velocities derived in Section \ref{sec:sq_ampl,surf_vel}, Equations (\ref{eq:sq_ampl}) and (\ref{eq:rms_surf_vel}). Given the assumptions (i)--(iv), one can show that the properties of an SBGW can be completely specified by a single function $S_h(\nu)$ (\citealt{Allen1999}; \citealt{Maggiore2008}), defined by
\begin{multline}
	\langle h_A^*(\nu,\bmath{n}) h_{A'}(\nu',\bmath{n}')\rangle\\
	=\delta(\nu-\nu')\delta^2(\bmath{n}-\bmath{n}')\delta_{AA'}\frac{S_h(\nu)}{8\pi}, \label{eq:def_Sh}
\end{multline}
where $\delta^2(\bmath{n}-\bmath{n}')\equiv\delta(\phi-\phi')\delta(\rmn{cos}\Theta-\rmn{cos}\Theta')$. Moreover, $h_A(\nu)$ are the Fourier amplitudes of the GW tensor $h_{ij}$ defined in Equation (\ref{eq:def_fourier_ampl}), with $A=\{+,\times\}$ labeling the two polarization states, and $\langle\rangle$ again denotes an ensemble average (i.e., the expectation value). The spectral density of the stochastic background $S_h(\nu)$ is a real, non-negative function satisfying $S_h(\nu)=S_h(-\nu)$ and it can be shown that the following relation holds:
\begin{equation}
	\Omega_{\rmn{GW}}(\nu)=\frac{4\pi^2}{3H_0^2}\nu^3S_h(\nu) \label{eq:Om_Sh}
\end{equation}
(cf., e.g., \citealt{Allen1999}; \citealt{Maggiore2008}).

In order to calculate the power spectral density $P_m(\omega)$ for an SBGW, we start with the correlation function $K_m^{t_0}(\tau)$,
\begin{eqnarray}
	\langle K_m^{t_0}(\tau)\rangle\mskip-15mu&=&\mskip-15mu\frac{64\pi^2}{225}\langle\ddot{h}_{ij}(t)\ddot{h}_{kl}^*(t')\rangle \left(\mathcal{Y}_{ij}^{2m}\right)^*\mathcal{Y}_{kl}^{2m}\nonumber\\
	&=&\mskip-15mu\frac{1024\pi^6}{225}\sum_{A,A'}\int_{-\infty}^{\infty} d\nu d\nu'\int d\Omega d\Omega'\nonumber\\
	&&\mskip-15mu\times\langle h_A(\nu,\bmath{n}) h_{A'}^*(\nu',\bmath{n}')\rangle e_{ij}^A(\bmath{n}) e_{kl}^{A'}(\bmath{n}')\nonumber\\
	&&\mskip-15mu\times\nu^2\nu'^2 e^{-i 2\pi\nu t}e^{i2\pi\nu' t'} \left(\mathcal{Y}_{ij}^{2m}\right)^*\mathcal{Y}_{kl}^{2m},\nonumber\\
	\label{eq:K_stoch1}
\end{eqnarray}
where in the first step Equations (\ref{eq:corrfunc}) and (\ref{eq:inv_expan_sph_harm}) were employed. Moreover, we set $t\equiv t_0-\tau/2$ and $t'=t_0+\tau/2$. In the second step, the plane wave expansion for the GW tensor of the SBGW in the long-wavelength approximation was used,
\begin{equation}
	h_{ij}(t)=\sum_{A}\int_{-\infty}^{\infty}d\nu\int d\Omega\,h_A(\nu,\bmath{n})e_{ij}^A(\bmath{n})e^{-i2\pi\nu t},
\end{equation}
cf. Equation (\ref{eq:pwe3}). Inserting Equation (\ref{eq:def_Sh}) into Equation (\ref{eq:K_stoch1}) and eliminating the delta functions by integrating over $\nu'$ and $\Omega'$ yields
\begin{eqnarray}
	\langle K_m^{t_0}(\tau)\rangle\mskip-15mu&=&\mskip-15mu\frac{128\pi^5}{225}\int_{-\infty}^{\infty} d\nu\,\nu^4 S_h(\nu)e^{i2\pi\nu \tau}\int d\Omega\nonumber\\
	&&\mskip-15mu\times\sum_{A}\sum_{i,j}\sum_{k,l}e_{ij}^A(\bmath{n}) e_{kl}^{A}(\bmath{n})\left(\mathcal{Y}_{ij}^{2m}\right)^*\mathcal{Y}_{kl}^{2m}.\nonumber\\
	\label{eq:K_stoch2}
\end{eqnarray}
Clearly, the right-hand side does not depend on $t_0$ and thus the correlation function solely depends upon time differences. Laborious calculations using Equations (\ref{eq:Y1})--(\ref{eq:n}) and (\ref{eq:e1})--(\ref{eq:v}) show that the second factor on the right-hand side of Equation (\ref{eq:K_stoch2}),
\begin{equation}
	c_m\equiv\int d\Omega\sum_{A}\sum_{i,j}\sum_{k,l}e_{ij}^A(\bmath{n}) e_{kl}^{A}(\bmath{n})\left(\mathcal{Y}_{ij}^{2m}\right)^*\mathcal{Y}_{kl}^{2m},
\end{equation}
is a constant that is independent of the azimuthal order $m$,
\begin{equation}
	c_m=6\;\;\;\forall\; m=\{0,\pm 1,\pm 2\}.
\end{equation}
It is interesting to note that for $m=0$ cross-polarization components do not contribute to the expectation value of the correlation function. However, in the case of $m=\{\pm 1,\pm 2\}$ the contributions of both polarization components are roughly of the same order.

Consequently, one is left with
\begin{equation}
	\langle K_m^{t_0}(\tau)\rangle=\frac{8}{75}\int_{-\infty}^{\infty} d\omega\,\omega^4 S_h(\omega)\,e^{i\omega\tau},
\end{equation}
which after Fourier-transforming in accordance with Equation (\ref{eq:psd}) yields the desired expression,
\begin{equation}
	P_m(\omega)=\frac{16\pi}{75}\omega^4 S_h(\omega), \label{eq:Pm_Sh}
\end{equation}
or, equivalently, expressed in terms of $\Omega_{\rmn{GW}}(\omega)$ using relation (\ref{eq:Om_Sh}),
\begin{equation}
	P_m(\omega)=\frac{32\pi^2}{25}\omega H_0^2\Omega_{\rmn{GW}}(\omega).\label{eq:Pm_Om}
\end{equation}
Equations (\ref{eq:Pm_Sh}) and (\ref{eq:Pm_Om}) nicely reflect the assumptions (i)--(iv) and
it should be noted that $P_m(\omega)$ is independent of the Hubble expansion rate $H_0$, so that experimental uncertainties related to $H_0$ do not enter the present formalism (cf. Equations (\ref{eq:Om}) and (\ref{eq:rho_c})).

However, in some cases it might be more convenient to have an expression available that directly relates $P_m(\omega)$ to the characteristic strain spectrum of an SBGW. Under the assumptions (i)--(iii), the characteristic strain spectrum is related to $\Omega_{\rmn{GW}}$ by the following equation \citep{Allen1997,Maggiore2000,DePies2007}:
\begin{equation}
	\Omega_{\rmn{GW}}(\omega)=\frac{1}{6H_0^2}\omega^2h_c(\omega)^2. \label{eq:om_hc}
\end{equation}
Consequently, when substituted we have
\begin{equation}
	P_m(\omega)=\frac{16\pi^2}{75}\omega^3 h_c(\omega)^2.
\end{equation}
Up to this point, the theory developed so far is still very general. In order to obtain numerical results for the mean-square amplitudes and rms surface velocities of normal modes excited by an SBGW, one now has to either specify a particular spectral energy density $\Omega_{\rmn{GW}}(\omega)$ or, equivalently, a particular characteristic strain spectrum $h_c(\omega)^2$, thereby focusing on a particular stochastic background. This is done in Section \ref{sec:results_sbgw1}.

Finally, using Equation (\ref{eq:Pm_Om}) we reformulate Equations (\ref{eq:sq_ampl}) and (\ref{eq:rms_surf_vel}) in terms of the spectral energy density $\Omega_{\rmn{GW}}(\omega)$ to obtain the general expressions in the case of an SBGW,
\begin{equation}
	\big\langle|A_N|^2\big\rangle=\frac{\pi^2}{25}\frac{R^8\chi_n^2}{\eta_N\omega_N I^2}H_0^2\Omega_{\rmn{GW}}(\omega_{N}), \label{eq:sq_ampl_stoch}
\end{equation}
and
\begin{equation}
	v_{N,s}=\left[\frac{\pi^2}{50}\frac{R^8\chi_n^2}{I^2}\frac{\omega_N} {\eta_N}H_0^2\Omega_{\rmn{GW}}(\omega_N)\Psi_N^2(R)\right]^\frac{1}{2}. \label{eq:rms_surf_vel_stoch}
\end{equation}
As in Section \ref{sec:GW_radiation1}, the numerical evaluation of these formulae for the mean-square amplitudes and rms surface velocities is postponed to the next section.

\section{Numerical results for the Sun}
\label{sec:results}

The theory developed in the preceding sections is very general and applies to any non-relativistic star and to arbitrary SBGWs or (quasi-)monochromatic GW sources. In this section, we focus on the Sun as our nearest (non-relativistic) star and present numerical results for resonance factors, mean-square amplitudes and rms surface velocities by making additional assumptions on the incoming gravitational radiation. We start with discussing the quantities that are directly related to or derived from the chosen solar model.

\subsection{Eigenfrequencies and eigenfunctions \label{sec:eigenmodes}}

In order to compute resonance factors, mean-square amplitudes, and surface velocities, knowledge of the eigenfrequencies $\omega_N$ and the radial as well as horizontal eigenfunctions, $\xi_{r,n2}(r)$ and $\xi_{h,n2}(r)$, of $l=2$ solar \textit{g}- and \textit{p}-mode oscillations is required. We note again that in the framework of the linearized theory of general relativity solely quadrupolar eigenmodes of non-relativistic stars can be excited by external GWs (cf. Section \ref{sec:sq_ampl,surf_vel}). Eigenfrequencies and eigenfunctions are computed using the standard Aarhus adiabatic oscillation package ADIPLS \citep{ChristensenDalsgaard2008} and the underlying solar model that we employ is Model S of \citet{ChristensenDalsgaard1996}, which uses the OPAL opacities and the OPAL equation of state. Eigenfunctions are normalized according to Equation (\ref{eq:normalization}).

\subsection{Damping rates \label{sec:damp}}

As can be seen from Equation (\ref{eq:sq_ampl}), or, alternatively, from Equations (\ref{eq:sq_ampl_pw}) and (\ref{eq:sq_ampl_stoch}), square amplitudes are the result of a balance between driving by external GWs and damping processes in the interior of the Sun and thus, damping rates are important quantities to be reliably specified. In the case of solar \textit{p} modes, we employ current experimental values obtained from data that has been recorded with the Michelson Doppler Imager (MDI) on board the \textit{SOHO} spacecraft (J. Schou, 2006, private communication). The crucial point, however, is the computation of \textit{g}-mode damping rates, since they cannot be obtained from observational data.

\begin{figure}[tb]
	\centering
	\includegraphics[width=0.47\textwidth]{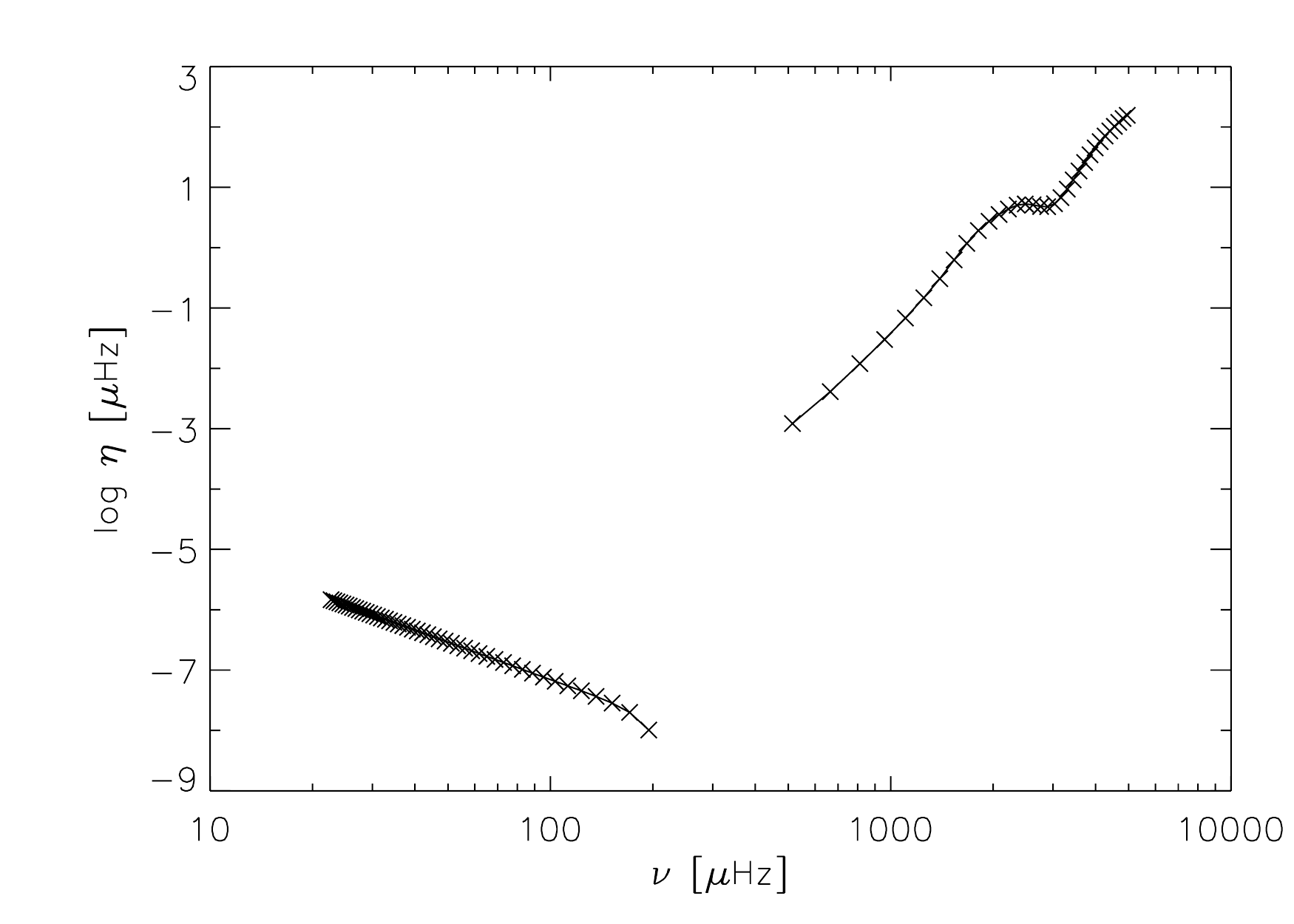}
	\caption{Combined damping rates of $l=2$ solar \textit{g} and \textit{p} modes (from left to right). Damping rates of \textit{g} modes are taken from \citet{Belkacem2009}; \textit{p} mode damping rates are extracted from MDI data (see the text for details).\label{fig:damp}}
\end{figure}

Damping rates for solar \textit{g} modes are taken from \citet{Belkacem2009}. These values have been computed with the non-adiabatic pulsation code MAD \citep{Dupret2002}, in which a time-dependent convection treatment presented and discussed in \citet{Grigahcene2005} is implemented. In Figure \ref{fig:damp}, the combined damping rates of solar \textit{p} and \textit{g} modes are depicted. Both data sets have been interpolated onto the adiabatic frequency grid computed with the ADIPLS code.

A few comments on \textit{g}-mode damping rates are in order. For $\nu< 110\,\mu\rmn{Hz}$, $\eta$ is a decreasing function of frequency and follows a very regular law, $\eta\propto \nu^{-3}$, which can also be recovered from quasi-adiabatic analytical computations (see \citealt{Belkacem2009}). This result is due to radiative damping in the inner core being the dominant contribution to the work integral and thus to the damping rates \citep{Belkacem2009}. However, for $\eta> 110\,\mu\rmn{Hz}$, time-dependent convection terms become significant and damping rates are sensitive to the parameter $\beta$ \citep{Grigahcene2005} in the perturbation of the energy closure equation. This parameter is used by \citet{Belkacem2009} to model convection--pulsation interactions and the value adopted by these authors was chosen such that good agreement between theoretical and observed damping rates of solar \textit{p} modes is achieved. However, there is no evidence that this value can readily be used in \textit{g}-mode calculations. Consequently, \citet{Belkacem2009} conclude that \textit{g}-mode damping rates for $\eta> 110\,\mu\rmn{Hz}$ are questionable. However, uncertainties related to the modeling of convection-oscillation interaction are not inherent in \textit{g}-mode damping rates below $110\,\mu\rmn{Hz}$.

\subsection{Long-wavelength approximation \label{sec:lwa}}

The long-wavelength approximation (see Section \ref{sec:theoretical_formulation}) if appropriate will apply to all GW sources, since it only requires the quotient $l_\rmn{GW}\equiv d/(\lambda/2\pi)=(2\pi d \nu)/c$ to be very small compared to unity. Here, $d$ is the diameter of the star under consideration, $c$ is the speed of light, $\lambda$ denotes the wavelength of the incident GW radiation, and $\nu$ its frequency. In the solar case one has $l_\rmn{GW}< 6\times 10^{-3}$ for the \textit{g}-mode frequencies ($\nu<200\,\mu\text{Hz}$) and $l_\rmn{GW}< 0.15$ for the \textit{p}-mode frequencies ($\nu<5\,\text{mHz}$) considered here and thus the long-wavelength approximation is well justified in the solar case. For stars more massive than the Sun radii increase whereas \textit{g}-mode oscillation frequencies decrease. Generally, the long-wavelength approximation is expected to be appropriate for \textit{g}-mode frequencies of all non-relativistic stars and it is also expected to be well motivated for \textit{p}-mode frequencies. However, the smallness of $l_\rmn{GW}$ has to be checked case by case.

\subsection{Resonance factors \label{sec:res_fac}}

In Section \ref{sec:sq_ampl,surf_vel}, we introduced the resonance factor,
\begin{equation}
	Q_{\rmn{res},N}=\frac{R^8\chi_n^2}{32\eta_{N}\omega_{N}^2 I^2}.
\end{equation}
It contains all quantities in the expression for the square amplitudes, Equation (\ref{eq:sq_ampl}), that are solely dependent on the solar model and that do not depend on the incoming GW radiation. Hence, the resonance factor associated with a particular normal mode provides a general measure for the ability of the Sun to resonate in this particular eigenmode, independent of the assumed external GW field.

\begin{figure}[tb]
	\centering
	\includegraphics[width=0.47\textwidth]{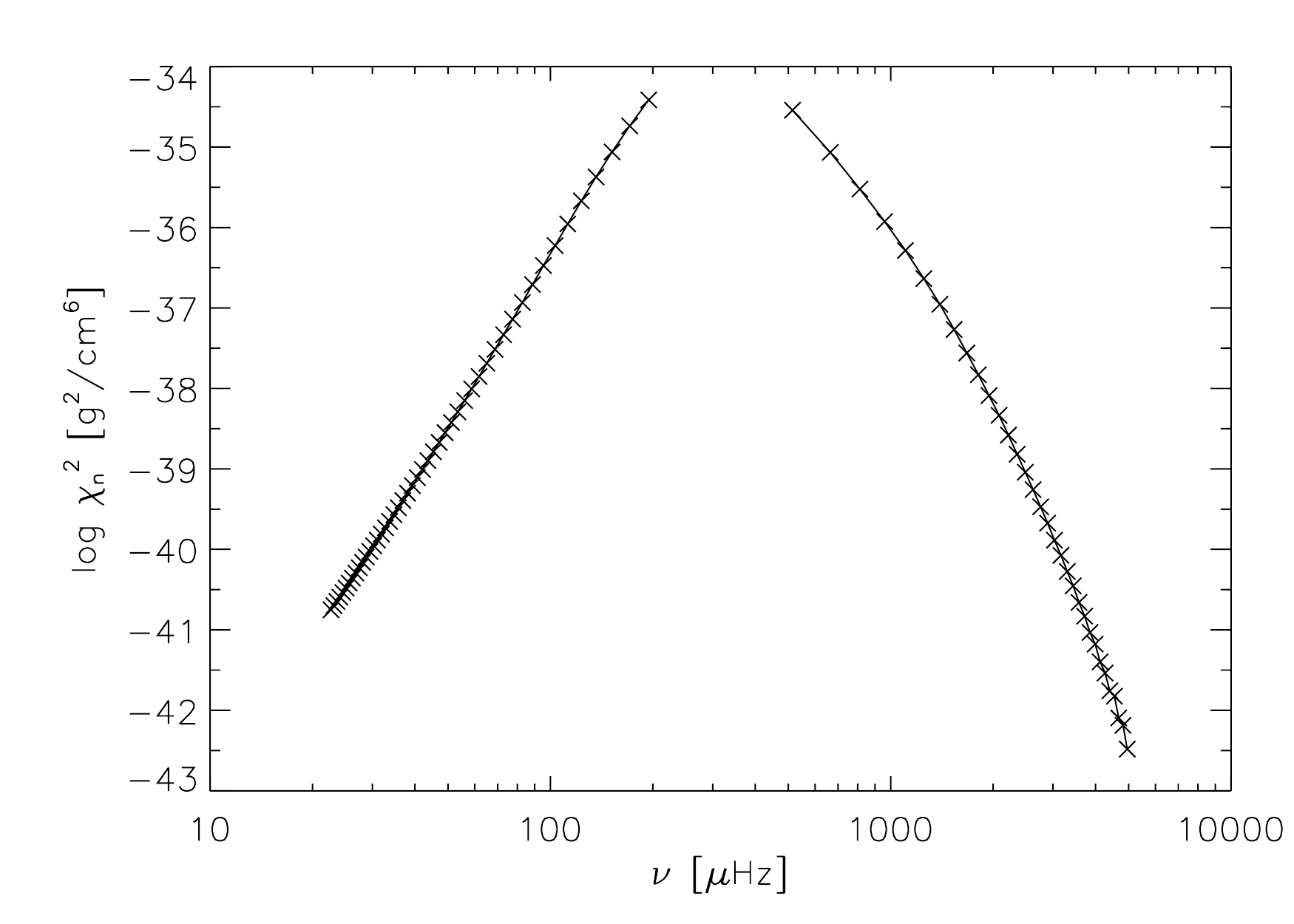} 
	\includegraphics[width=0.47\textwidth]{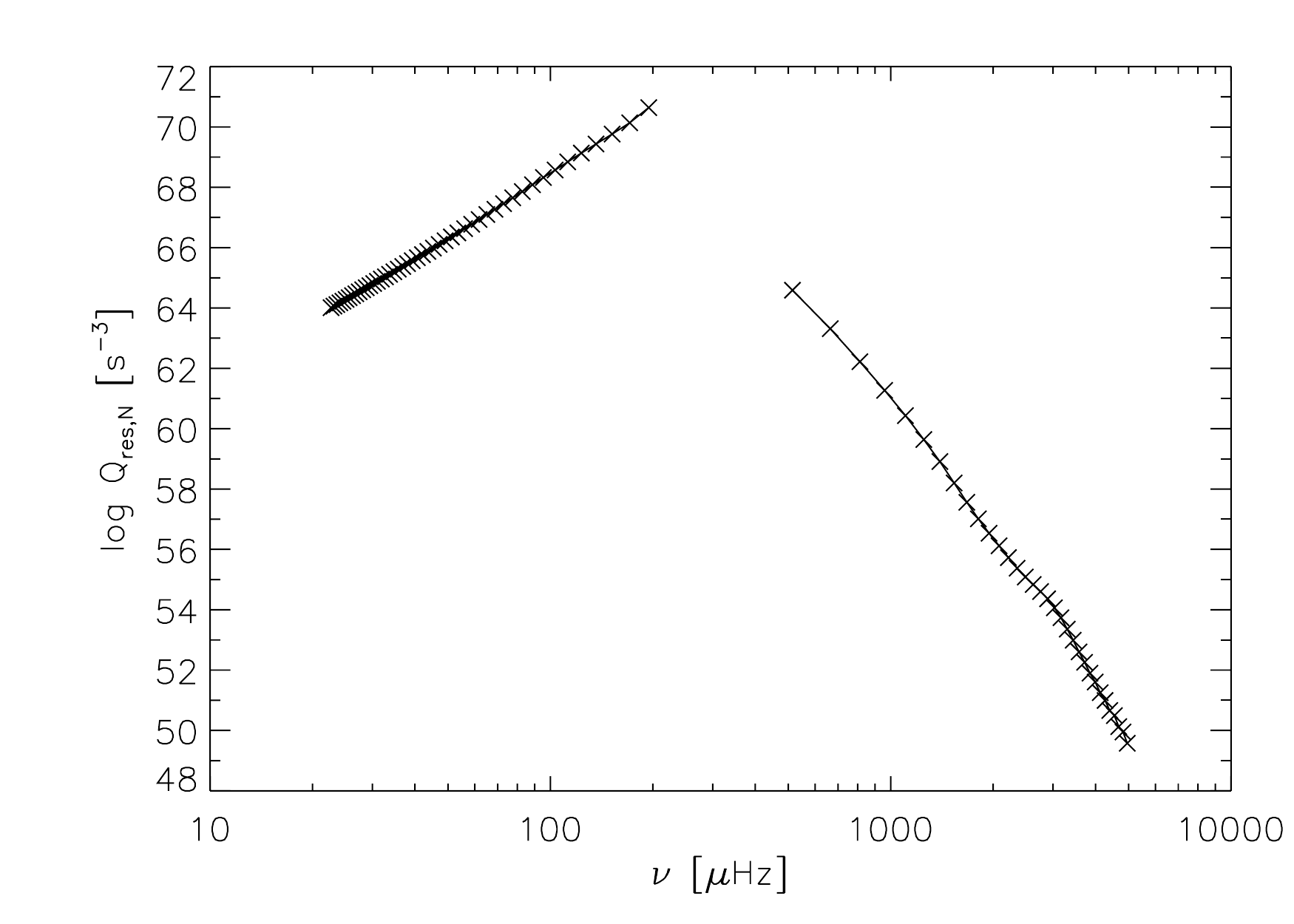} 
	\caption{Numerical values for $\chi^2_n$ (top) and for resonance factors (bottom) of $l=2$ solar \textit{g} and \textit{p} modes (from left to right). \textit{g} modes show a much higher general ability for resonance in the case of external forcing by gravitational waves.\label{fig:chi-res_fac}}
\end{figure}

Numerical results for the resonance factors $Q_{\rmn{res},N}$ based on the computations described in Sections \ref{sec:eigenmodes} and \ref{sec:damp} are depicted in Figure \ref{fig:chi-res_fac}. In addition, we present numerical values for the quantity $\chi_n^2$, which is defined in Equation (\ref{eq:chi}) and which is among the main contributors to the resonance factor. We note that the high values for the resonant factors present in Figure \ref{fig:chi-res_fac} result from the normalization of the eigenfunctions (cf. Section \ref{sec:sq_ampl,surf_vel}). Only relative values have a direct meaning and are to be considered.

From Figure \ref{fig:chi-res_fac}, it is evident that the overall spectral shape of $Q_{\rmn{res},N}$ is dominantly determined by $\chi_n^2$, with the offset between the \textit{g}-mode and the \textit{p}-mode branch being due to division by damping rates and squared frequency. While a power law would only be a crude first-order estimate for \textit{p} modes, it is interesting to note that the \textit{g}-mode resonance factors very regularly follow a power law. The latter fact is due to the power-law behavior of the damping rates in the \textit{g}-mode frequency range (cf. Section \ref{sec:damp}), together with the power law for \textit{g}-mode frequencies present in the numerical results for the quantity $\chi_n^2$ (top plot of Figure \ref{fig:chi-res_fac}).

As is also obvious from Figure \ref{fig:chi-res_fac}, \textit{g} modes show up to seven orders of magnitude higher resonance factors than \textit{p} modes and thus \textit{g} modes have a much higher general ability to resonate in the case of this excitation mechanism. However, in order to be able to decide whether or not solar \textit{g} modes are indeed excited more intensively than \textit{p} modes, one has to compare the corresponding mean-square amplitudes. This is addressed in Sections \ref{sec:results_as2} and \ref{sec:results_sbgw2}.

\subsection{Excitation by distinct astrophysical sources \label{sec:results_as}}

In this section, we present numerical results for the mean-square amplitude and rms surface velocity of a particular solar normal mode that is excited by a quasi-monochromatic and resolvable astrophysical source, by which we mean that the number of sources in the same, sufficiently wide frequency bin does not exceed one, or, at least, the GW signals of other sources in this bin should be comparatively small, such that they can be neglected. Apart from an SBGW of cosmological origin (see Section \ref{sec:results_sbgw}), ultra-compact binary systems are among the most relevant astrophysical sources for gravitational radiation in the low-frequency band of stellar oscillations to be ``observed'' from the solar system. In particular, AM CVn stars, which are interacting ultra-compact binaries (white dwarfs accreting helium-rich material from a compact companion), are the only \textit{known} sources in this frequency range (see Section \ref{sec:results_as1}), albeit many other astrophysical sources like massive black hole mergers, compact objects spiraling into massive black holes, etc., are expected. In view of this it seems advisable to explicitly evaluate the still very general formalism presented in Section \ref{sec:GW_radiation1} for the case of a binary system. In this section, we are only interested in the quasi-monochromatic radiation associated with binaries before the final spiral-in phase, that is, the timescale for frequency shift due to loss of energy by gravitational radiation is large compared to the GW period. We review the expected intrinsic strain amplitudes for known and expected systems. This allows us to deduce a realistic order of magnitude estimate for the expected strain amplitudes to be employed in calculations for square amplitudes and surface velocities; the results of these computations are presented in Section \ref{sec:results_as2}.

\subsubsection{Assumptions on the incoming gravitational radiation \label{sec:results_as1}}

According to the quadrupole formula, gravitational radiation in TT gauge from a binary system in circular orbit is given by (\citealt{Timpano2006}; \citealt{Maggiore2008})
\begin{equation}
	h_{ij}(t)=A_+ e^{-i2\pi \nu_{\rmn{gw}}t} e_{ij}^+(\bmath{e}_z)+A_\times e^{-i2\pi\nu_{\rmn{gw}}t}e_{ij}^\times (\bmath{e_z}), \label{eq:Bi_wf1}
\end{equation} 
where the polarization amplitudes are
\begin{multline}
	A_+=\frac{4}{rc^4}(G\mathcal{M})^{5/3}(\pi\nu_{\rmn{gw}})^{2/3}\\
	\times\frac{1+\cos^2\Theta}{2}e^{i2\pi \nu_{\rmn{gw}}r/c},
\end{multline}
\begin{multline}	
	A_-=i\frac{4}{rc^4}(G\mathcal{M})^{5/3}(\pi\nu_{\rmn{gw}})^{2/3}\\
	\times\cos\Theta e^{i2\pi \nu_{\rmn{gw}}r/c}.\label{eq:Bi_wf3}
\end{multline}
Here $\mathcal{M}=(M_1M_2)^{3/5}/(M_1+M_2)^{1/5}$ is the chirp mass, with $M_1$ and $M_2$ being the component masses, $\Theta$ is the inclination of the binary, $r$ its distance, and $\nu_{\rmn{gw}}$ denotes twice the instantaneous orbital frequency. Moreover, we already assumed that the radiation is coming from the $-\bmath{e}_z$ direction and we also adopted the convention introduced in Section \ref{sec:GW_radiation1} of allowing complex quantities and taking the real part at the end of the calculation. We note that adopting this convention allows us to make the waveforms of a binary system directly compatible to the form of gravitational radiation assumed in Section \ref{sec:GW_radiation1}. Moreover, we note that galactic double white dwarf (DWD) systems, DWD systems being the most common binary systems, are expected to have invariably circular orbits \citep{Willems2007,Nelemans2009,Ruiter2010}. In addition, the assumption of a circular orbit is also very well satisfied in the case of galactic AM CVn stars, which are also commonly expected to be circular (\citealt{Nelemans2004}; \citealt{Roelofs2006}; \citealt{Roelofs2007a}; \citealt{Ruiter2010}).

Following \citet{Timpano2006}, \citet{Roelofs2006}, and \citet{Roelofs2007a}, we define the GW strain amplitude $h$ by
\begin{equation}
	h=\left[\frac{1}{2}\left(|A_+|^2+|A_-|^2\right)\right]^{1/2}.
\end{equation}
Furthermore, we take the mean power spectral densities $\overline{P_{\pm 2}}(\omega_N)$ of $N=\{n,2,\pm 2\}$ modes as a measure for predicting the effect of such individual astrophysical sources on solar oscillations, such that the corresponding mean-square amplitudes and rms surface velocities (cf. Equations (\ref{eq:sq_ampl_pw}) and (\ref{eq:surf_vel_pw})) become
\begin{equation}
\overline{|A_N|^2}=\frac{\pi^2}{15}\frac{R^8\chi_n^2\omega_N^2}{\eta_N I^2}h^2, \label{eq:sq_ampl_as}
\end{equation}
and
\begin{equation}
	\overline{v_{N,s}}=\left[\frac{\pi^2}{30}\frac{R^8\chi_n^2\omega_N^4}{\eta_N I^2}h^2\Psi_N^2(R)\right]^\frac{1}{2}, \label{eq:rms_surf_vel_as}
\end{equation}
with the bar over $P_{\pm 2}(\omega_N)$, $|A_N|^2$, and $v_{N,s}$ indicating that the mean of the $N=\{n,2,\pm 2\}$ modes is considered.

\begin{table}[tb]
	\begin{center}
	\caption{Predictied Gravitational Wave Strain Amplitudes $h$ and Frequencies $\nu_{\rmn{gw}}$ of Five AM CVn Binary Systems \citep{Roelofs2007a} \label{tab:1}}
	\begin{tabular}{lcc}%{@{}lcc}
		\tableline\tableline
		Star 		& $\nu_{\rmn{gw}}$		& $h$ \\
					& (mHz)					&	\\
		\tableline
		AM CVn		& $1.944$	& $2.0^{+0.4}_{-0.3}\times 10^{-22}$ \\ 
		HP Lib		& $1.813$	& $3.7^{+0.6}_{-0.8}\times 10^{-22}$ \\
		CR Boo		& $1.360$	& $2.1^{+0.4}_{-0.5}\times 10^{-22}$ \\
		V803 Cen	& $1.253$	& $3.0^{+0.5}_{-0.7}\times 10^{-22}$ \\
		GP Com		& $0.7158$	& $[4.0-6.6]\times 10^{-23}$ \\
		\tableline
	\end{tabular}
	\end{center}
\end{table}

The only \textit{known}, guaranteed GW sources in the low-frequency band that have well-determined strain amplitudes with reliable error bars are five AM CVn systems, which are listed in Table \ref{tab:1} \citep{Roelofs2007a}. These sources have strain amplitudes of the order of $h\sim 10^{-22}$. In addition, there are several tens of known DWD binaries ($\nu_{\rmn{gw}}\sim 7\times 10^{-7}-4\times 10^{-4}\,\rmn{Hz}$; cf. \citealt{Nelemans2005}) and AM CVn systems ($\nu_{\rmn{gw}}\sim 5\times 10^{-4}- 6\times 10^{-3}\,\rmn{Hz}$)\footnote{A list of known AM CVn stars, their properties, and corresponding references is available at: \url{http://www.astro.ru.nl/$\sim$nelemans/dokuwiki/\\doku.php?id$=$verification\_binaries:am\_cvn\_stars}}, their strain amplitudes (cf., e.g., \citealt{Nelemans2004}), however, can only be estimated for lack of accurate distance measurements. Moreover, focusing on frequencies $\nu_{\rmn{gw}}\sim 10^{-3}- 10^{-2}\,\rmn{Hz}$, more than 10,000 galactic DWD binaries and thousand to more than 10,000 AM CVn systems are expected in the strain amplitude range of $h\sim 10^{-24}-10^{-22}$ (\citealt{Nelemans2004}; \citealt{Roelofs2007b}; \citealt{Ruiter2010}).

Besides these known sources, massive black hole binaries in the mass range $10^{3}-10^{8}\,M_\odot$ are expected to provide even stronger gravitational radiation at $\mu\rmn{Hz}$ and mHz frequencies. For example, a representative circular massive black hole binary with masses $M_2=0.1M_1=10^{5}\,M_\odot$ at redshift $z=1$ produces a strain amplitude of $h\sim 10^{-20}$ at $\nu_{\rmn{gw}}\sim 10^{-5}\,\rmn{Hz}$ before the final spiral-in phase \citep{Sesana2005}.

The preceding discussion shows that for an order of magnitude estimate for amplitudes and surface velocities of solar normal modes excited by a particular quasi-monochromatic astrophysical source, it is reasonable to assume the source to have maximal strain amplitudes of $h\sim 10^{-20}$.

\subsubsection{Numerical Results \label{sec:results_as2}}

\begin{figure}[tb]
	\centering
	\includegraphics[width=0.47\textwidth]{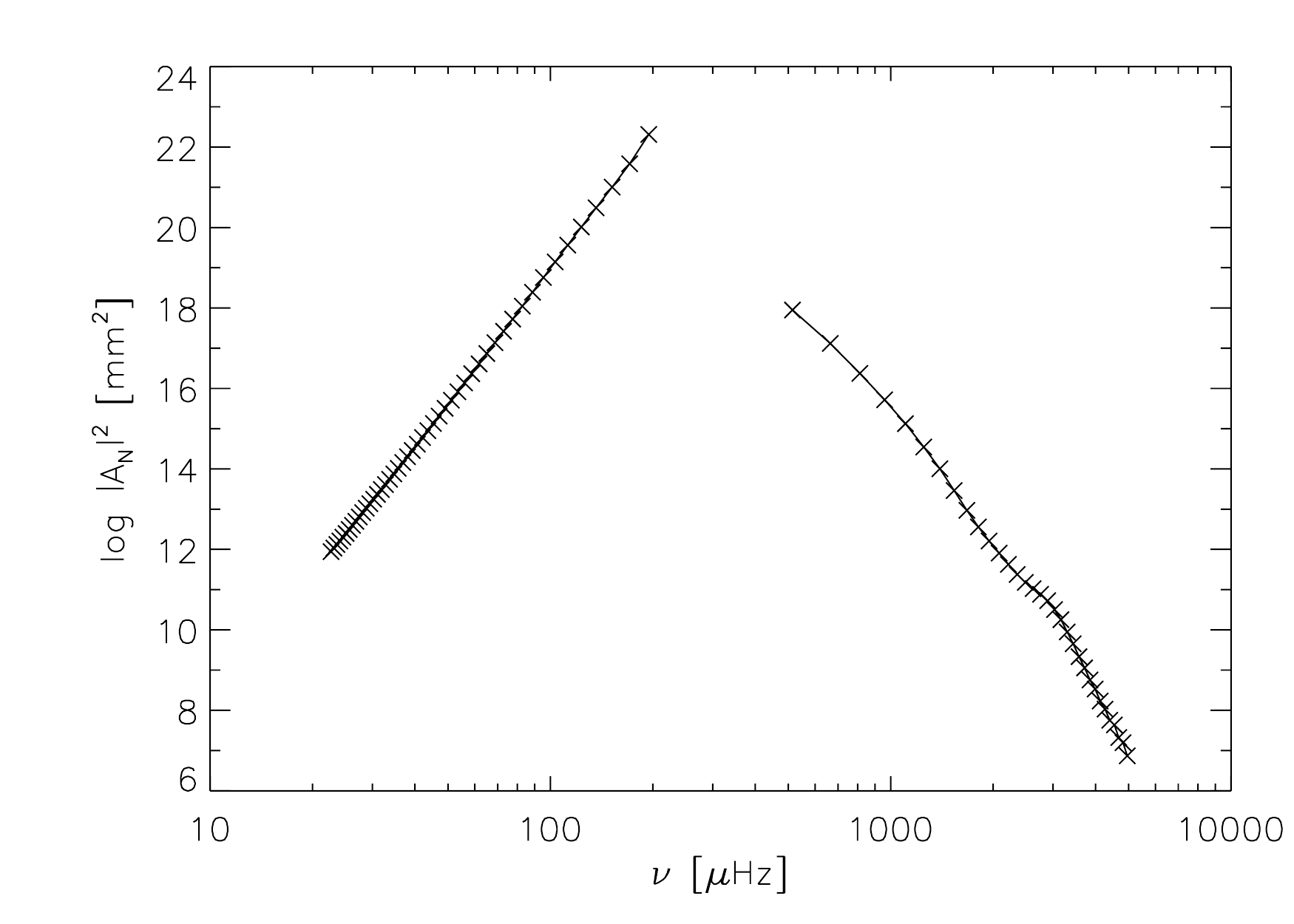}
	\includegraphics[width=0.47\textwidth]{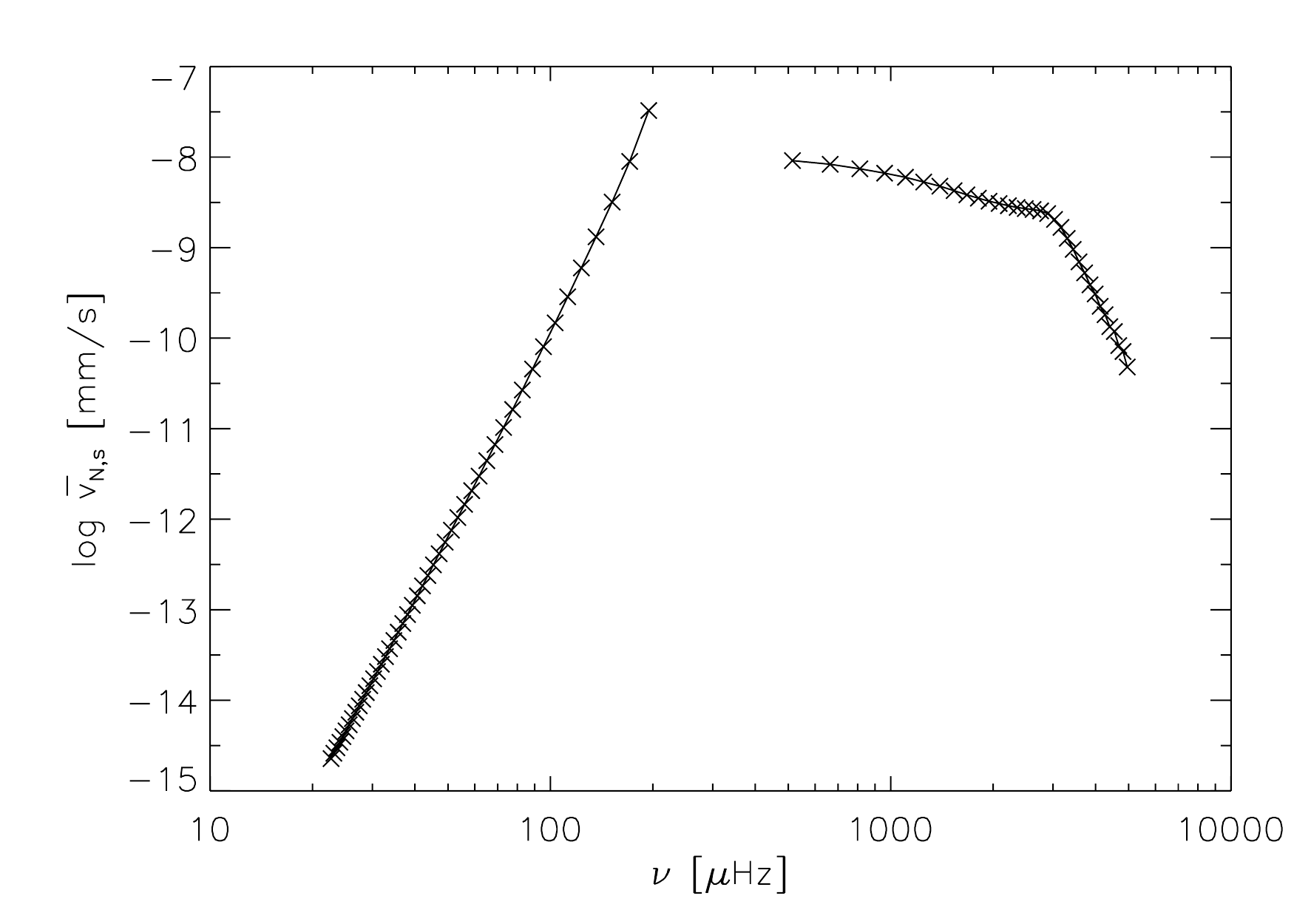}
	\caption{Numerical results for mean-square amplitudes (top) and rms surface velocities (bottom) of $l=2$ solar \textit{g} and \textit{p} modes (from left to right), assuming a constant gravitational wave strain amplitude of $h=1\times 10^{-20}$ over frequency from individual astrophysical sources as discussed in Sections \ref{sec:GW_radiation1} and \ref{sec:results_as1}.\label{fig:results_as}}
\end{figure}

In order to provide an order of magnitude estimate for the effect of external forcing by quasi-monochromatic GW sources on solar oscillations, we assume that for each eigenfrequency of the Sun there is a GW source at the very same frequency with a gravitational strain amplitude of $h=1\times 10^{-20}$ (cf. Section \ref{sec:results_as1}). Using the damping rates discussed in Section \ref{sec:damp}, eigenfunctions and eigenfrequencies as well as quantities directly related to the solar model, the former and the latter being discussed in Section \ref{sec:eigenmodes}, and Equations (\ref{eq:sq_ampl_as}) and (\ref{eq:rms_surf_vel_as}), we obtain the results depicted in Figure \ref{fig:results_as}.

Again we note that in the plot showing the mean-square amplitudes, only relative values have a direct meaning, since the high absolute values are an immediate consequence of the adopted normalization condition for the eigenfunctions (cf. Section \ref{sec:sq_ampl,surf_vel}).

As is obvious from Figure \ref{fig:results_as}, low-frequency \textit{g} modes show smaller mean-square amplitudes than low-frequency \textit{p} modes. This is due to suppression by the $\omega^4$ dependence on frequency present in the power spectral density $P_m(\omega)$ (cf. Equation (\ref{eq:Pm_pw}) and the results for the resonance factors in Figure \ref{fig:chi-res_fac}). However, mean-square amplitudes of high-frequency \textit{g} modes ($\nu_{\rmn{gw}}>80\,\mu\rmn{Hz}$) exceed the corresponding values of low-frequency \textit{p} modes by up to five orders of magnitude. This shows that, due to the definition of the square amplitudes, Equation (\ref{eq:vel_expan}), and due to the normalization condition (\ref{eq:normalization}), high-frequency \textit{g} modes are more strongly excited than \textit{p} modes by several orders of magnitude.

\begin{figure}[tb]
	\centering
	\includegraphics[width=0.47\textwidth]{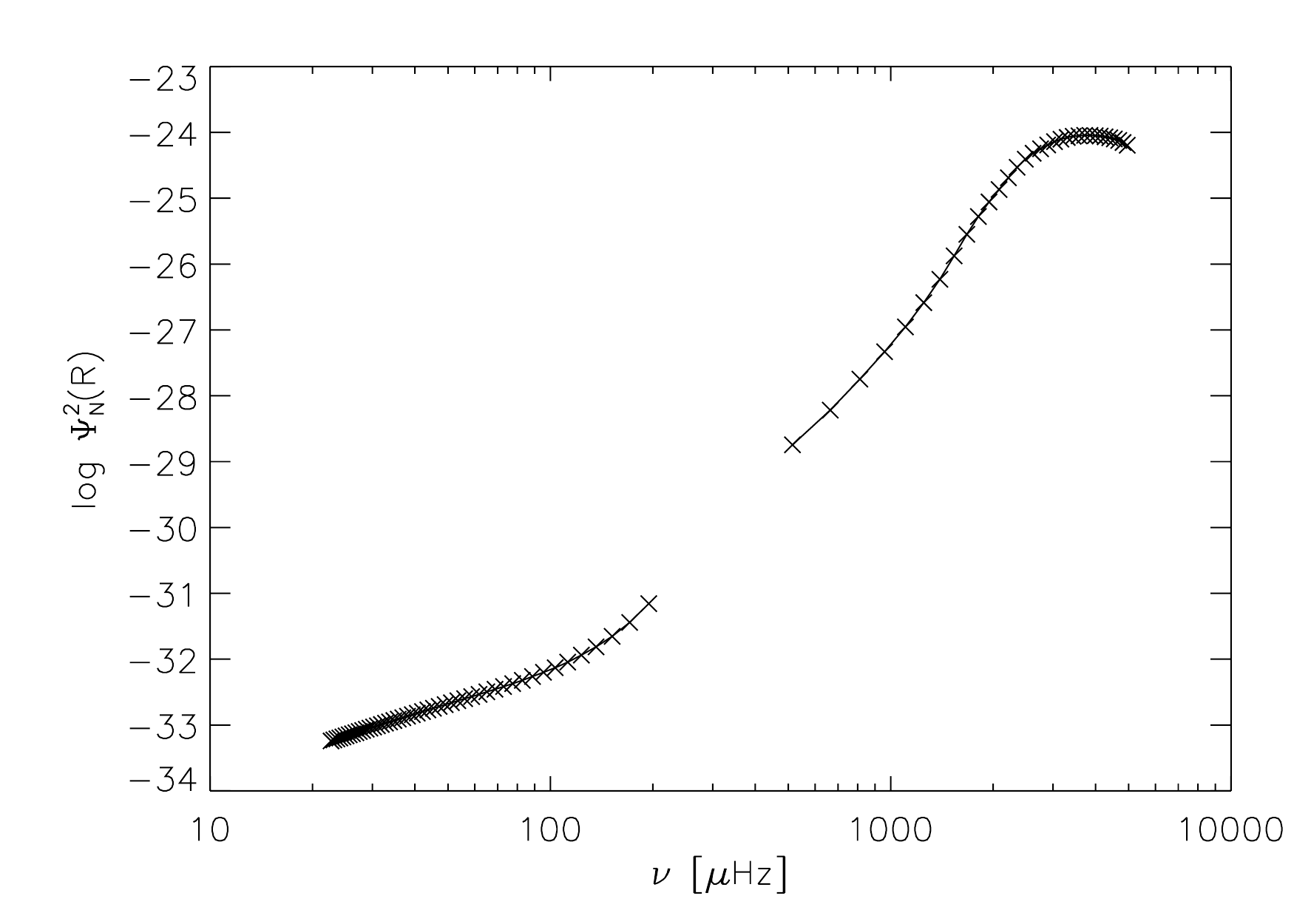}
	\caption{Numerical values for the quantity $\Psi_N^2(R)$ (cf. Equation (\ref{eq:psi})) of $l=2$ solar \textit{g} and \textit{p} modes (from left to right); \textit{g}-mode values are orders of magnitude smaller than \textit{p}-mode values. A power-law behavior for asymptotic \textit{g} modes is clearly visible.\label{fig:psi}}
\end{figure}

In Figure \ref{fig:results_as}, we also show numerical results for rms surface velocities, which are found to reach values of $\overline{v_{N,s}}\la 10^{-8}\,\rmn{mm}\,\rmn{s}^{-1}$. With perhaps a few exceptions, \textit{g}-mode values generally lie below \textit{p}-mode values. Consequently, although high-frequency \textit{g} modes are excited more strongly than \textit{p} modes, this behavior is not obvious any more at the surface. This is because of \textit{g} modes being exponentially damped in the convection zone. In the analytical expressions for the rms surface velocities, Equations (\ref{eq:rms_surf_vel}), (\ref{eq:surf_vel_pw}), and (\ref{eq:rms_surf_vel_as}), damping of \textit{g} modes in the convection zone can be seen to be modeled by the quantity $\Psi_N^2(R)$, which is defined in Equation (\ref{eq:psi}) and which essentially measures the amplitude of the eigenfunctions at the surface of the star. Numerical results for this quantity are depicted in Figure \ref{fig:psi}, in which damping of \textit{g} modes in the convection zone is nicely reflected by \textit{g}-mode values being orders of magnitude smaller than \textit{p}-mode values.

We finally note that both mean-square amplitudes and rms surface velocities for \textit{g} modes show very regular power-law behavior. This is due to the power-law dependence on frequency shown by the damping rates (cf. Section \ref{sec:damp}), the quantity $\chi_n^2$ (cf. Section \ref{sec:res_fac}), and the power spectral density $P_m$ (cf. Equation (\ref{eq:Pm_pw})). In the case of surface velocities, it is important to note that the quantity $\Psi_N^2(R)$ also shows a power law for asymptotic (low-frequency) \textit{g} modes. The deviation from power-law behavior inherent in rms surface velocity values at high \textit{g}-mode frequencies is predominantly due to a departure from power-law behavior of $\Psi_N^2(R)$ at frequencies above 100\,Hz, which is evident from Figure \ref{fig:psi}. Assessing the significance of the surface velocities found in this subsection by comparing them to corresponding values in the case of stochastic excitation by turbulent convection (first two terms on the right-hand side of Equation (\ref{eq:EOM1})) is postponed to Section \ref{sec:comp}.

\subsection{Excitation by a stochastic background of gravitational waves}
\label{sec:results_sbgw}
In this section, we discuss numerical results for mean-square amplitudes and rms surface velocities of solar normal modes excited by an SBGW. As is pointed out in Section \ref{sec:GW_radiation2}, conducting such calculations requires to either specify a particular spectral energy density $\Omega_{\rmn{gw}}(\omega)$ or, equivalently, a characteristic strain spectrum $h_c(\omega)^2$ for the SBGW to be expected. Therefore, we start with briefly reviewing possible sources for an SBGW in order to deduce a realistic assumption on the spectral energy density to be employed.

\subsubsection{Assumptions on the incoming gravitational radiation \label{sec:results_sbgw1}}

A (rather weak) contribution to the SBGW at $\mu\rmn{Hz}$ frequencies is expected from astrophysical sources. The characteristic strain spectrum of an SBGW originating from coalescing massive and supermassive black hole binaries at the center of galaxies is expected to peak at frequencies $\nu_{\rmn{gw}}\sim 10^{-10}\,\rmn{Hz}$, with $h_c^2\sim 10^{-28}$ (\citealt{Sesana2004}; \citealt{Enoki2004}); at $\nu_{\rmn{gw}}\sim 10^{-4}\,\rmn{Hz}$ these authors find $h_c^2\sim 10^{-38}$, which translates into $\Omega_{\rmn{GW}}\sim 10^{-10}$ using Equation (\ref{eq:om_hc}). The upper bound on the entire frequency range of $10^{-12}\,\rmn{Hz}\la\nu_{\rmn{gw}}\la 10^{-1}\,\rmn{Hz}$ is $\Omega_{\rmn{GW}}\sim 10^{-8}$, which is reached at $\nu_{\rmn{gw}}\sim 10^{-6}\,\rmn{Hz}$. A background due to galactic and extragalactic detached white dwarf binaries is generally weaker than the one by black hole binaries and has been calculated by \citet{Nelemans2001} and \citet{Farmer2003}, respectively. In the case of galactic binaries, the former authors find $\Omega_{\rmn{GW}}\sim 10^{-11}$ at $\nu_{\rmn{gw}}\sim 10^{-4}\,\rmn{Hz}$, with an upper bound of $\Omega_{\rmn{GW}}\sim 10^{-10}$ on the whole frequency range, $10^{-4}\,\rmn{Hz}\la\nu_{\rmn{gw}}\la 10^{-3}\,\rmn{Hz}$, whereas the latter authors in the case of extragalactic binaries obtain $\Omega_{\rmn{GW}}\sim 10^{-13}- 10^{-12}$ at $\nu_{\rmn{gw}}\sim 10^{-4}\,\rmn{Hz}$, with an upper limit of $\Omega_{\rmn{GW}}\sim 10^{-11}$ on the entire frequency range, $10^{-6}\,\rmn{Hz}\la\nu_{\rmn{gw}}\la 10^{-1}\,\rmn{Hz}$.

The second general category of an SBGW is formed by cosmological sources. A weak contribution at $\mu\rmn{Hz}$ frequencies is expected from an SBGW that originates from quantum fluctuations during inflation (\citealt{Starobinskii1979}; \citealt{Allen1988}; \citealt{Sahni1990}; \citealt{Bar-Kana1994}; \citealt{Kuroyanagi2009}). After a decrease with frequency as $\nu_{\rmn{gw}}\sim 10^{-2}$ from $10^{-18}\,\rmn{Hz}$ to $10^{-16}\,\rmn{Hz}$, typical inflation models predict almost scale-invariant spectra of GWs over many orders of magnitude in frequency, e.g., $10^{-16}\,\rmn{Hz}\la\nu_{\rmn{gw}}\la 10^{-1}\,\rmn{Hz}$, up to a region near a cutoff frequency, which is related to the end of reheating. The constant spectral energy density typically found in this quasi-scale-invariant frequency regime is $\Omega_{\rmn{GW}}\la 10^{-14}$. An SBGW arising from amplification of vacuum fluctuations in string theory inspired pre-big-bang models is thought to be very weak in the low-frequency range of stellar \textit{g}-mode pulsations but it would be among candidates for a dominant SBGW at high frequencies, $\nu_{\rmn{gw}}> 1\,\rmn{Hz}$. Such models generally yield a spectrum that shows an increase at low frequencies proportional to $\nu^3$ up to a turn-over frequency $\nu_s$, above which it is found that $\Omega_{\rmn{GW}}\sim \nu^{3-2\mu}$, and finally the spectrum cuts off at a frequency $\nu_1$, which is expected to be around $10^{10}\,\rmn{Hz}$ (\citealt{Brustein1995}; \citealt{Buonanno1997}; \citealt{Mandic2006}). Depending on the model parameters, $\nu_s$, $\mu$, and $\nu_1$, these models are able to produce spectral energy densities at high frequencies, e.g., $\nu_{\rmn{gw}}>10\,\rmn{Hz}$, that lie close to present observational bounds on an SBGW. However, at frequencies of stellar \textit{g}-mode pulsations below $10^{-3}\,\rmn{Hz}$ the spectral energy density $\Omega_{\rmn{GW}}$ is thought to be smaller than the SBGW predicted by the inflation models discussed above by orders of magnitude.

Cosmic strings might be the dominant source for an SBGW in the low-frequency range of solar and stellar \textit{g}-mode pulsations and they presently seem to be the only possible source to produce an SBGW that lies close to current observational bounds at low frequencies as well as at high frequencies. Cosmic strings are one-dimensional topological defects that can be formed during symmetry breaking phase transitions in the early universe \citep{Kibble1976}. In addition, string-theory-inspired inflation scenarios also produce cosmic strings, which are referred to as cosmic superstrings (\citealt{Jones2002}; \citealt{Sarangi2002,Dvali2004}; \citealt{Jones2003,Copeland2004}). Whereas in the former case of field theoretic strings the reconnection probability is unity, $p=1$, cosmic superstrings have a reconnection probability less than unity, with expected values of $10^{-3}<p<1$ \citep{Jackson2005}. A network of cosmic (super)strings gives rise to an SBGW via incoherent superposition of bursts from cusps and kinks on these strings. This mechanism of GW production has been studied intensively in the past (e.g., \citealt{Vilenkin1981,Caldwell1992}; \citealt{Caldwell1996}; \citealt{Damour2001,Damour2005}; \citealt{Siemens2007,Oelmez2010}). After a decrease or a steep rise at the low-frequency end of the spectrum, depending on the assumed mechanism that determines the string length, gravitational backreaction (``small loops'') or large-scale dynamics of the network (``large loops''), the spectral energy density $\Omega_{\rmn{GW}}$ typically shows a scale-invariant behavior over many orders of magnitude. Approximate analytical calculations by \citet{Oelmez2010} show that this constant is solely determined by two model parameters, the reconnection probability $p$ and the dimensionless constant $G\mu$ that characterizes the gravitational interaction of strings with string tension $\mu$; $G$ denotes Newton's gravitational constant. The current observational bound on the string tension from cosmic microwave background (CMB) observations is $G\mu<6.1\times 10^{-7}$ (\citealt{Pogosian2004}; \citealt{Jeong2005}). In the small loop case, the aforementioned constant is given by
\begin{equation}
	\Omega_{\rmn{GW}}\propto \frac{G\mu}{p},
\end{equation}
whereas in the case of large loops one finds
\begin{equation}
	\Omega_{\rmn{GW}}\propto \frac{\sqrt{G\mu}}{p}
\end{equation}
\citep{Oelmez2010}.

In view of the spectral properties of various possible candidates for an SBGW mentioned above and the observational bounds to be listed below, we adopt the following scenarios in order to estimate mean-square amplitudes and rms surface velocities of solar normal modes, in particular the \textit{g} modes, excited by an SBGW. We focus on an SBGW due to cosmic strings and neglect the contributions of all other possible sources, since their contributions to an SBGW at solar \textit{g}- and \textit{p}-mode pulsation frequencies are expected to be very small compared to possible spectra produced by cosmic strings. Furthermore, we consider the following two cases. First, we adopt an optimistic scenario and assume a constant spectral energy density over the frequency range of solar \textit{g}- and low-frequency \textit{p}-mode oscillations of $\Omega_{\rmn{GW}}(\omega)=1\times 10^{-5}$. This is approximately the case at solar \textit{g}-mode frequencies for a small loop model of radiation from cusps with parameters $p=5\times 10^{-3}$, $G\mu=10^{-7}$, and $\epsilon\sim 10^{-7}$ (cf. \citealt{Siemens2007}; \citealt{Oelmez2010}). We note that this model is perfectly compatible with all observational bounds, such as the current pulsar limit at low frequencies around $10^{-8}\,\rmn{Hz}$ \citep{Jenet2006}, the direct bound around 100\,Hz deduced from the Laser Interferometer Gravitational-wave Observatory (LIGO) science run S5 \citep{Abbott2009}, the bound based on CMB and matter spectra \citep{Smith2006}, the indirect bound due to the big bang nucleosynthesis (BBN; \citealt{Cyburt2005}), and the CMB bound on the string tension mentioned above. Second, we consider a rather pessimistic case and assume a constant spectral energy density over the frequency range of solar \textit{g}- and low-frequency \textit{p}-mode oscillations of $\Omega_{\rmn{GW}}(\omega)=1\times 10^{-8}$. Such a value for the constant part of the GW spectrum would, for example, result from a large loop model of superstrings with $G\mu=10^{-11}$ and $p=10^{-1}$ or from a large loop model of field theoretic stings ($p=1$) with $G\mu=10^{-9}$ \citep{Oelmez2010}. Both of these models include radiation from cusps and kinks, they already reach the scale-invariant regime at lower frequencies than the \textit{g}-mode frequency range, and both are perfectly consistent with all current observational bounds \citep{Oelmez2010}. For the remainder of this paper, we employ $H_0=73\,\rmn{km}\,\rmn{s}^{-1}\rmn{Mpc}^{-1}$ for the Hubble parameter, as is done by \citet{Siemens2007} and \citet{Oelmez2010}. We note, however, that our results are independent of the adopted value of $H_0$ and its experimental uncertainty (cf. Section \ref{sec:GW_radiation2}).

\subsubsection{Numerical Results \label{sec:results_sbgw2}}

\begin{figure}[tb!]
	\centering
	\includegraphics[width=0.47\textwidth]{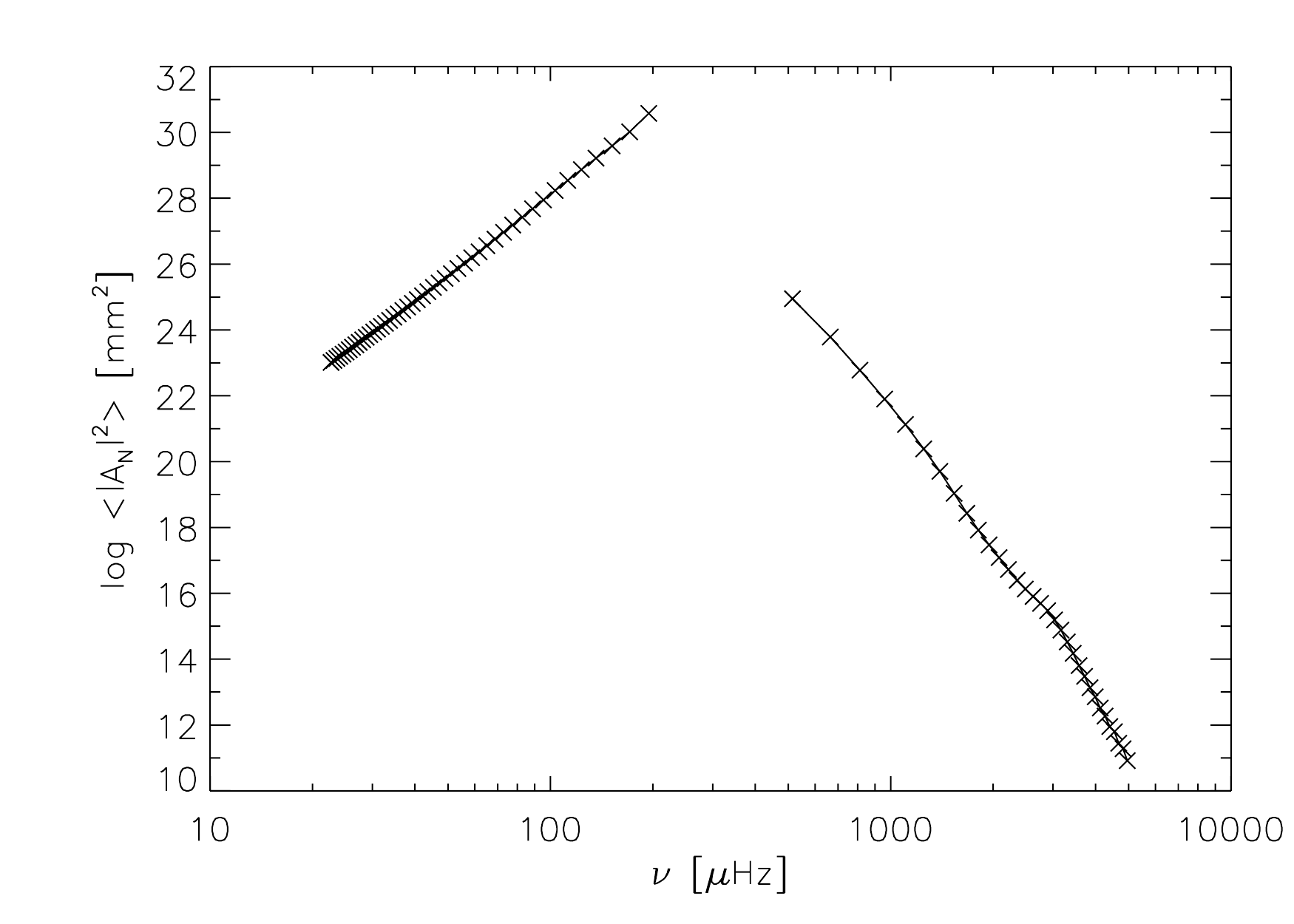}
	\includegraphics[width=0.47\textwidth]{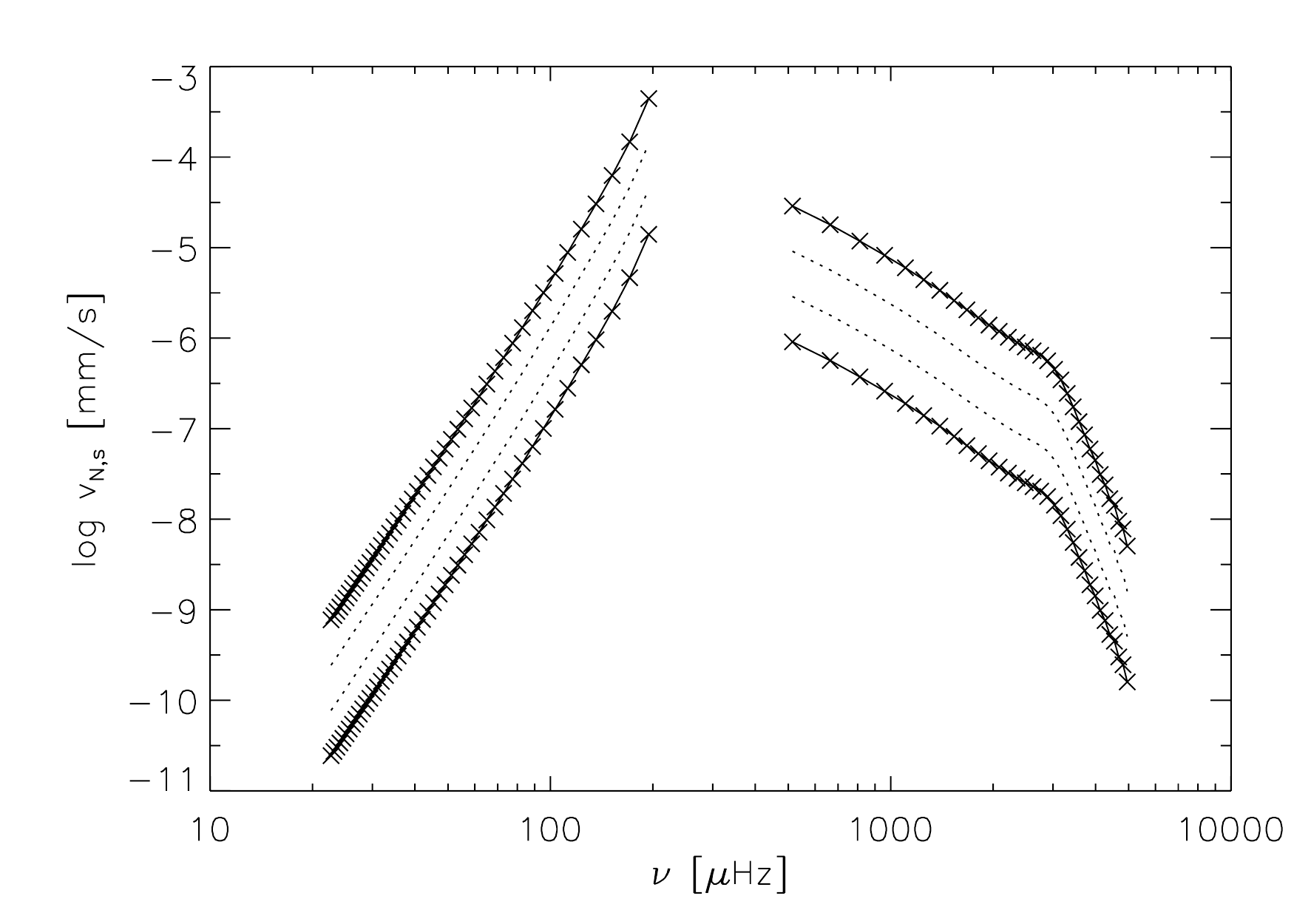}
	\caption{Top: numerical results for mean-square amplitudes of $l=2$ solar \textit{g} and \textit{p} modes (from left to right) assuming $\Omega_{\rmn{GW}}(\nu)$ to be constant, $\Omega_{\rmn{GW}}=1\times 10^{-5}$. High-frequency (low-radial order) \textit{g} modes are more strongly excited than low-frequency \textit{p} modes by up to six orders of magnitude. Bottom: numerical results for rms surface velocities of the same modes assuming a constant spectral energy density of $\Omega_{\rmn{GW}}=1\times 10^{-5}$ (top curve) and $\Omega_{\rmn{GW}}=1\times 10^{-8}$ (bottom curve). The dotted curves in between correspond to setting $\Omega_{\rmn{GW}}=1\times 10^{-6}$ and $\Omega_{\rmn{GW}}=1\times 10^{-7}$, respectively.\label{fig:results_stoch}}
\end{figure}

In Figure \ref{fig:results_stoch}, we present numerical results for mean-square amplitudes and rms surface velocities of quadrupolar solar \textit{g} and \textit{p} modes excited by an SBGW. The calculations have been performed using Equations (\ref{eq:sq_ampl_stoch}) and (\ref{eq:rms_surf_vel_stoch}), eigenfunctions and eigenfrequencies as well as quantities directly related to the solar model as discussed in Section \ref{sec:eigenmodes}, damping rates as presented in Section \ref{sec:damp}, and the model scenarios for an SBGW due to cosmic strings described in Section \ref{sec:results_sbgw1}.

As depicted in the top plot of Figure \ref{fig:results_stoch}, mean-square amplitudes of \textit{g} modes above $40\,\mu\rmn{Hz}$ exceed corresponding values for low-frequency \textit{p} modes by up to six orders of magnitude and thus we conclude that these \textit{g} modes are much more strongly excited than \textit{p} modes (cf. the definition of square amplitudes, Equation (\ref{eq:vel_expan}), and the normalization condition, Equation (\ref{eq:normalization})). The reason for the very low-frequency (high-radial order) \textit{g} modes having smaller mean-square amplitudes than low-frequency \textit{p} modes is suppression of low-frequency modes that results from an explicit frequency dependence of the power spectral density $P_m(\omega)$, $P_m(\omega)\propto \omega$ (cf. Equation (\ref{eq:Pm_Om}) and the results for the resonance factors presented in Section (\ref{sec:res_fac})). Note that in the case of excitation by a particular astrophysical source, $P_m(\omega)$ even includes a $\omega^4$ suppression factor for low-frequency modes (see Equation (\ref{eq:Pm_pw})) and as a result of this, mean-square amplitudes of \textit{g} modes below $80\,\mu\rmn{Hz}$ lie below corresponding values for low-frequency \textit{p} modes (see Section \ref{sec:results_as2}). As a rather marginal remark we again mention that in the top plot of Figure \ref{fig:results_stoch}, only relative values have a direct meaning, since absolute values are strongly dependent on the actual normalization constant $I$, cf. Equation (\ref{eq:normalization}). Therefore, if $\Omega_{\rmn{GW}}(\omega)$ is considered to be constant, the actual value of $\Omega_{\rmn{GW}}$ is irrelevant to the discussion of mean-square amplitudes and we arbitrarily set $\Omega_{\rmn{GW}}=1\times 10^{-5}$ in this plot.

Again, and for the same reason as stated in Section \ref{sec:results_as2}, mean-square amplitudes of \textit{g} modes and rms surface velocities of asymptotic (low-frequency) \textit{g} modes show characteristic power-law behavior. The deviation from a power law visible in the rms surface velocities of high-frequency \textit{g} modes is predominantly due to the non-power-law behavior of the quantity $\Psi_N^2$ at frequencies above 100\,Hz (cf. Figure \ref{fig:psi} and Section \ref{sec:results_as2}).

In Figure \ref{fig:results_stoch}, we plot numerical results for rms surface velocities adopting the optimistic scenario (upper curve, $\Omega_{\rmn{GW}}=1\times 10^{-5}$ in the considered frequency range) and the rather pessimistic scenario (bottom curve, $\Omega_{\rmn{GW}}=1\times 10^{-8}$ in the considered frequency range), which are discussed in Section \ref{sec:results_sbgw1}. The dotted curves in between are plotted to guide the eye and correspond to setting $\Omega_{\rmn{GW}}=1\times 10^{-6}$ and $\Omega_{\rmn{GW}}=1\times 10^{-7}$. As in the case of excitation by a particular astrophysical source (Section \ref{sec:results_as2}), we again find that most of the \textit{g}-mode values lie below the \textit{p}-mode rms surface velocities, which is due to the quantity $\Psi_N^2(R)$ that in the present formalism can be considered as to model the exponential damping of \textit{g} modes in the convection zone. Setting $\Omega_{\rmn{GW}}=1\times 10^{-5}$ we obtain \textit{g}-mode rms surface velocities in the range $10^{-9}\,\rmn{mm}\,\rmn{s}^{-1}\la v_{N,s}\la 10^{-3}\,\rmn{mm}\,\rmn{s}^{-1}$, whereas for $\Omega_{\rmn{GW}}=1\times 10^{-8}$ values are found to be of the order of $10^{-11}\,\rmn{mm}\,\rmn{s}^{-1}\la v_{N,s}\la 10^{-5}\,\rmn{mm}\,\rmn{s}^{-1}$, with increasing values toward higher frequencies. In this context, it is important to stress that rms surface velocities scale with the square root of $\Omega_{\rmn{GW}}(\omega)$, cf. Equation (\ref{eq:rms_surf_vel_stoch}), and thus, for example, shifting $\Omega_{\rmn{GW}}$ by two orders of magnitude only changes surface velocities by one order of magnitude. In the following section, we will focus on a comparison with present estimates for \textit{g}-mode rms surface velocities in the case of stochastic excitation by turbulent convection, thereby assessing the significance of the results presented here.

\subsection{Comparison with stochastic excitation by turbulent convection \label{sec:comp}}

Theoretical quantitative estimates for rms surface velocities of low-degree ($l=1,2,3$) \textit{g} modes that are stochastically excited by turbulent convection differ from each other by orders of magnitude. The current range for quadrupolar \textit{g}-mode rms surface velocities is $10^{-3}\,\rmn{mm}\,\rmn{s}^{-1}\la v_{N,s}\la 1\,\rmn{mm}\,\rmn{s}^{-1}$ (for a review on the topic see, e.g., \citealt{Appourchaux2010}). These order of magnitude differences are predominantly due to the choice of the assumed turbulent eddy-time correlation function. Using a Lorentzian profile, \citet{Belkacem2009} find rms surface velocities of the order of $v_{N,s}\sim 0.1- 1\,\rmn{mm}\,\rmn{s}^{-1}$ for quadrupolar \textit{g} modes, whereas employing a Gaussian profile these authors essentially reproduce the results of \citet{Kumar1996} ($10^{-3}\,\rmn{mm}\,\rmn{s}^{-1}\la v_{N,s}\la 0.1\,\rmn{mm}\,\rmn{s}^{-1}$), who also used a Gaussian function. A rather crude estimate of \textit{g}-mode amplitudes using the equipartition ansatz of \citet{Goldreich1977} and without referring to an explicit mode excitation and damping model was reported by \citet{Gough1985}, who obtained rms surface velocities for quadrupolar \textit{g} modes of the order of $v_{N,s}\sim 0.1\,\rmn{mm}\,\rmn{s}^{-1}$.

Comparing these findings with our theoretical estimates for rms surface velocities of quadrupolar \textit{g} modes excited by GWs of a particular astrophysical source of $v_{N,s}\la 10^{-8}\,\rmn{mm}\,\rmn{s}^{-1}$ (see Section \ref{sec:results_as2}), we conclude that this excitation mechanism seems not to be relevant to the excitation of quadrupolar \textit{g} modes. However, this picture changes dramatically if we consider excitation by an SBGW. In this case and depending on the adopted scenario (optimistic: $\Omega_\rmn{GW}=1\times 10^{-5}$ in the \textit{g}-mode frequency range; pessimistic: $\Omega_\rmn{GW}=1\times 10^{-8}$ in the \textit{g}-mode frequency range; see the discussion in Section \ref{sec:results_sbgw2}), we find maximal surface velocities in the range of $10^{-5}-10^{-3}\,\rmn{mm}\,\rmn{s}^{-1}$ (cf. Figure \ref{fig:results_stoch}). Consequently, present models for an SBGW due to cosmic strings are able to excite solar \textit{g} modes up to values that lie very close to (or possibly even within) the presently expected range for rms surface velocities of solar \textit{g} modes in the case of excitation by turbulent convection. Thus, we conclude that once the experimental sensitivity will allow for the detection of individual quadrupolar \textit{g} modes in the near future, and provided that rms surface velocities due to excitation by turbulent convection in fact lie in the presently predicted range, it might be possible to measure the effect of an SBGW on quadrupolar solar \textit{g} modes, or, at least, it might be possible to impose additional constraints on an SBGW.

\section{Conclusions}
\label{sec:conclusion}

In the present paper, a hydrodynamic formalism has been developed that describes excitation of normal modes of non-relativistic stars by externally generated GWs. In particular, we focused on stationary GW signals and we considered excitation by individual astrophysical sources, such as binary systems, as well as excitation by an SBGW of astrophysical or cosmological origin.

The equation of motion that governs the internal motions of a star under external forcing by GWs is derived from the field equations of general relativity by applying the linearized theory approach, which is valid by definition of non-relativistic stars, by adopting the long-wavelength approximation, which is valid for the Sun and which is expected to be valid for all non-relativistic stars (cf. Section \ref{sec:lwa}), and by assuming non-relativistic internal motions of the star. This Euler-like equation is then transformed into an equation of motion solely for the oscillations of the star by following the procedure discussed by \citet{Samadi2001a}, thereby assuming adiabatic oscillations that are decoupled from the turbulent medium and incompressible turbulence. However, in contrast to \citet{Samadi2001a} we did not adopt the Cowling approximation and we showed that terms related to perturbations in the Newtonian gravitational potential can be retained as far as oscillations are concerned. As a by-product of applying this procedure, we obtain a generalized and unified theoretical description of the mode excitation processes due to turbulent convection and GWs, combined in a single consistent forced wave equation. Focusing solely on excitation by GWs, we were able to analytically solve the resulting inhomogeneous linear differential equation, with the general solution given by an expression for the mean-square amplitudes of stellar normal modes, Equation (\ref{eq:sq_ampl_gen}). Only quadrupolar ($l=2$) normal modes are excited. In the case of stationary GW signals, we were able to further simplify the aforementioned expression and, as the main result of the first part of this paper, we obtain comparatively simple relations for the mean-square amplitudes and rms surface velocities of stellar normal modes, Equations (\ref{eq:sq_ampl}) and (\ref{eq:rms_surf_vel}). In Section \ref{sec:GW_radiation}, we derived analytical expressions for the power spectral density $P_m(\omega)$ of the GW tensor's spherical accelerations $\ddot{h}_m$ in the case of a particular astrophysical source of GW radiation and in the case of an SBGW.

Building on these theoretical results and focusing on the Sun as our nearest (non-relativistic) star, we computed and presented numerical results for mean-square amplitudes and rms surface velocities of quadrupolar solar normal modes excited by the aforementioned types of GW radiation. We find a very regular power-law behavior for resonance factors, mean-square amplitudes, and rms surface velocities of \textit{g} modes. As is shown by the numerical results for the resonance factors, solar \textit{g} modes have a much higher general ability to be excited than \textit{p} modes and indeed, as shown by the results for mean-square amplitudes, \textit{g} modes with frequencies above $80\,\mu\rmn{Hz}$ (excitation by a particular astrophysical source) or $40\,\mu\rmn{Hz}$ (excitation by an SBGW) are excited more intensively than \textit{p} modes by many orders of magnitude. Suppression of low-frequency (high-radial order) \textit{g} modes results from an explicit power-law frequency dependence of the power spectral density $P_m(\omega)$ ($\propto\omega^4$ and $\propto\omega$ in the above mentioned cases, respectively). 

In the case of excitation by individually resolvable astrophysical sources, we obtain maximal rms surface velocities of quadrupolar solar \textit{g} modes of $v_{N,s}\sim 10^{-8}\,\rmn{mm}\,\rmn{s}^{-1}$, assuming an intrinsic GW strain amplitude of a binary system of $h=1\times 10^{-20}$, which is a realistic (albeit optimistic) assumption at $\mu\rmn{Hz}$ frequencies as argued in Section \ref{sec:results_as1}. Compared to corresponding results for excitation by turbulent convection, this excitation mechanism due to individual astrophysical GW sources yields values that are four orders of magnitude smaller and thus it does not seem to be relevant to the excitation of solar \textit{g} modes, provided that the current range for quadrupolar solar \textit{g}-mode rms surface velocities that results from various predictions by different theoretical models is correct (at present: $10^{-3}\,\rmn{mm}\,\rmn{s}^{-1}\la v_{N,s}\la 1\,\rmn{mm}\,\rmn{s}^{-1}$; see the discussion in Section \ref{sec:comp}). However, this picture changes dramatically when considering an SBGW. As argued in Section \ref{sec:results_sbgw1}, cosmic (super)strings presently seem to be the only source of GW radiation at low frequencies that is likely to produce an SBGW close to current detection limits. Adopting an optimistic scenario of $\Omega_{\rmn{GW}}=1\times 10^{-5}$ and a rather pessimistic scenario of $\Omega_{\rmn{GW}}=1\times 10^{-8}$ in the solar \textit{g}-mode frequency range, we obtain maximal rms surface velocities of $10^{-5}-10^{-3}\,\rmn{mm}\,\rmn{s}^{-1}$. Maximal rms surface velocity values thus lie very close or might possibly even lie within the expected amplitude range for quadrupolar solar \textit{g} modes excited by turbulent convection, which is currently believed to be responsible for \textit{g}-mode excitation (e.g., \citealt{Appourchaux2010}). We note that both models for the SBGW are in perfect agreement with all observational bounds on an SBGW to date (see Section \ref{sec:results_sbgw1}). We also note that when shifting $\Omega_{\rmn{GW}}$ by several orders of magnitude, rms surface velocities are shifted only by half the amount of orders of magnitude, since the latter depend on $\Omega_{\rmn{GW}}$ only through the square root of $\Omega_{\rmn{GW}}$.

Hence, we conclude that once it will be possible to detect individual solar \textit{g} modes in the near future, one will either be able to measure the effect of an SBGW due to cosmic strings on solar \textit{g}-mode oscillations, thereby detecting GWs and using the Sun as a detector for GWs, or one will be able to deduce additional constraints on an SBGW at $\mu\rmn{Hz}$ and mHz frequencies. We repeat that GWs are only able to excite quadrupolar ($l=2$) oscillation modes and that turbulent convection excites all modes stochastically. Hence, GWs can be detected if $l=2$ mode amplitudes show a significantly different behavior than it would be expected from excitation by turbulent convection only; in particular, the ideal case is dominance of GW excitation, that is, $l\ne 2$ mode amplitudes are surmounted by amplitudes of $l=2$ modes. If this is not the case, one can deduce an upper bound on an SBGW by employing Equations (\ref{eq:sq_ampl_stoch}) and (\ref{eq:rms_surf_vel_stoch}) and assuming that excitation of $l=2$ oscillation modes is entirely due to excitation by GWs. Even if it turns out that the former case (the Sun as a detector for GWs) does not apply, the latter case would still be very interesting, given the fact that to date there is no method and experiment to derive a direct and significant upper bound on an SBGW at intermediate frequencies between the pulsar bound at low frequencies around $10^{-8}\,\rmn{Hz}$ and the bounds by Earth-based interferometric detectors at high frequencies around 100\,Hz. This intermediate frequency region would be particularly interesting in the context of cosmic string models for an SBGW, since such a bound would significantly constrain possible values for $\epsilon$ in small loop models with high values of $\Omega_{\rmn{GW}}$ for the constant part of the spectrum and therefore it would in particular constrain the parameter space of small loop models, i.e., models where the size of string loops is determined by gravitational backreaction.

In future work, we will apply the method developed in this paper to stars other than the Sun focusing in particular on those stars, for which, due to a different stellar structure, the excitation mechanism proposed in the present paper might be even more effective compared to excitation by turbulent convection. For stars more massive than the Sun and having larger radii $R$, oscillation amplitudes will benefit from the effective $R^2$-dependence present in the formulae for mean-square amplitudes and square surface velocities (cf. Equations (\ref{eq:sq_ampl_gen}), (\ref{eq:sq_ampl}), and (\ref{eq:rms_surf_vel})). However, making definite further statements about other stars in this regard is intricate, since detailed knowledge about the eigenfunctions of a star is needed to calculate the resonance factors, which determine the strength of mean-square amplitudes and rms surface velocities. In combination with experimental data from the \textit{Kepler} mission, the application of the method developed in this paper to stars other than the Sun might enable us to either use stars as giant hydrodynamic detectors for GWs, and this would set the scene for the detection of GWs by asteroseismology, or it might enable us to deduce a significant direct upper bound on an SBGW at intermediate frequencies.


%% Included in this acknowledgments section are examples of the
%% AASTeX hypertext markup commands. Use \url without the optional [HREF]
%% argument when you want to print the url directly in the text. Otherwise,
%% use either \url or \anchor, with the HREF as the first argument and the
%% text to be printed in the second.

\acknowledgments

The authors are grateful to M.-A. Dupret and K. Belkacem for sharing their \textit{g}-mode damping rates published in \citet{Belkacem2009} and to J. Christensen-Dalsgaard for the provision of the Aarhus adiabatic pulsation package ADIPLS. The authors also wish to thank O. von der L\"uhe for valuable discussions. M.R. acknowledges support from the European Helio- and Asteroseismology Network (HELAS), which was funded as Coordination Action under the European Commission's Framework Programme 6. D.M.S. acknowledges support from the German National Academic Foundation.



\appendix

\section[]{Derivation of the Euler equations}
\label{sec:app_1}

This appendix is concerned with explicitly deriving the equation of continuity (\ref{eq:continuity}) and the equation of motion (\ref{eq:Euler1}) for the internal velocity field from the general relativistic conservation equation of energy and momentum (\ref{eq:EMC2}). 

As described in Sections \ref{sec:theoretical_formulation} and \ref{sec:EOM}, the following assumptions and approximations are in order: 
\begin{enumerate}
	\item[(i)] \textit{Long-wavelength approximation.} The GW tensor $h_{\mu\nu}$ and hence by Equation (\ref{eq:Riemann}) the Riemann tensor have no spatial dependence over the volume of the star.
	\item[(ii)] \textit{TT gauge.} The GW tensor can be assumed to be in TT gauge (cf. Equation (\ref{eq:tt_gauge})) when computing components of the Riemann tensor due to the invariance property of the Riemann tensor.
	\item[(iii)] \textit{Newtonian internal motions.} Components of the internal velocity field $\bmath{v}$ of the star are small compared to the speed of light, $v^{i}\ll c$.
	\item[(iv)] \textit{Linearization.} Only terms linear in $h_{\mu\nu}$ need be retained (linearized theory assumption); in addition, only terms linear in the coordinates $x^\mu$ are retained.
\end{enumerate}
Furthermore, we define the following two tensors for convenience:
\begin{equation}
	T^{\mu\nu}=\rho u^\mu u^\nu+\left[p\left(\frac{u^\mu u^\nu}{c^2}+g^{\mu\nu}\right)\right]\equiv M^{\mu\nu}+P^{\mu\nu}.
\end{equation}
The components of the four-velocity field are given by $u^\mu=\gamma(c,\bmath{v})$, where $\gamma\equiv 1/\sqrt{1+\bmath{v}^2/c^2}$.

\subsection{Equation of continuity}

The equation of continuity (\ref{eq:continuity}) follows from the zero-component of the general relativistic conservation equation of energy and momentum,
\begin{equation}
	\partial_{\mu}M^{\mu 0}+\partial_{\mu}P^{\mu 0}=-\Gamma_{00}^0 T^{00}-\Gamma_{\nu 0}^\nu T^{00}, \label{eq:zero_EMC}
\end{equation}
where we only retained terms including $T^{00}$ on the right-hand side due to (iii).

Using $\gamma\simeq 1$ in accordance with (iii), we calculate
\begin{eqnarray}
	\partial_{\mu}M^{\mu 0}&=&\partial_{0}(\gamma \rho c^2)+\partial_i(\gamma^2\rho v^i c)\nonumber\\
	&\simeq& c\left[\frac{\partial\rho}{\partial t}+\bmath{\nabla\cdot}(\rho\bmath{v})\right]. \label{eq:M1}
\end{eqnarray}
Furthermore, according to (i)--(iv) and Equations (\ref{eq:FNC1})--(\ref{eq:FNC3}) we have
\begin{eqnarray}
	\partial_\mu P^{\mu 0}&=&(\partial_0 p)\left(\gamma^2+g^{00}\right)+(\partial_i p)\left(\gamma^2\frac{v^i}{c}+g^{i0}\right)\nonumber\\
	&&\mskip10mu+p\left[\partial_0\gamma^2+\partial_i\left(\gamma^2\frac{v^i}{c}\right)+\partial_\mu g^{\mu0}\right]\nonumber\\
	&\simeq& p\partial_\mu g^{\mu0}\nonumber\\
	&\simeq& 0,\label{eq:P1}
\end{eqnarray}
since $g^{00}\simeq -1$, $g^{0i}\simeq 0$, and
\begin{equation}
	\partial_\mu g^{\mu 0}\simeq -\frac{4}{3} \left(R^{0\phantom{l}i}_{\phantom{0}k\phantom{i}i}+R^{0\phantom{i}i}_{\phantom{0}i\phantom{i}k}\right)x^k=0.
\end{equation}
The first term in brackets vanishes, $R^{0\phantom{l}i}_{\phantom{0}k\phantom{i}i}=0$, as can be immediately seen by employing Equation (\ref{eq:Riemann}). The second term in brackets is also readily seen to be zero by using Equation (\ref{eq:Riemann}) and applying (ii), $R^{0\phantom{i}i}_{\phantom{0}i\phantom{i}k}=0$.

The Christoffel symbols
\begin{equation}
	\Gamma_{\mu\nu}^{\rho}=\frac{1}{2}g^{\rho\sigma}\left(\partial_\mu g_{\nu\sigma}+\partial_\nu g_{\mu\sigma}-\partial_\sigma g_{\mu\nu}\right)
\end{equation}
are evaluated in Fermi normal coordinates by using expressions (\ref{eq:FNC1})-(\ref{eq:FNC3}). The symbols needed in Equation (\ref{eq:zero_EMC}) are easily obtained as follows:
\begin{equation}
	\Gamma_{00}^{0}=\frac{1}{2}g^{00}\partial_0 g_{00}+g^{0i}(\partial_0 g_{0i}-\frac{1}{2}\partial_i g_{00})\simeq 0. \label{eq:Gamma0}
\end{equation}
This is due to $g^{0i}$ and $\partial_0 g_{00}$ already being of second order in the coordinates: $g^{0i}\simeq 0$ and 
\begin{equation}
	\partial_0 g_{00}\simeq -R_{0k0l}\partial_0(x^kx^l)=0. \label{eq:d0g00}
\end{equation}
Furthermore,
\begin{eqnarray}
	\Gamma_{\nu0}^{\nu}&=&\frac{1}{2}g^{\nu\sigma}\partial_0 g_{\nu\sigma}\nonumber\\
	&=&\frac{1}{2}g^{00}\partial_0 g_{00}+g^{0i}\partial_0 g_{0i}+\frac{1}{2}g^{ij}\partial_0 g_{ij}\nonumber \\
	&\simeq& 0, \label{eq:Gammanu}
\end{eqnarray}
where we additionally used
\begin{equation}
\partial_0 g_{ij}\simeq-\frac{1}{3}R_{ikjl}\partial_0(x^kx^l)=0.
\end{equation}
Inserting expressions (\ref{eq:M1}), (\ref{eq:P1}), (\ref{eq:Gamma0}), and (\ref{eq:Gammanu}) into Equation (\ref{eq:zero_EMC}) finally yields the desired result,
\begin{equation}
	\frac{\partial\rho}{\partial t}+\bmath{\nabla\cdot}(\rho\bmath{v})=0, \label{eq:continuity_ap}
\end{equation}
which is the usual equation of continuity.

\subsection{The Euler equation}

The equation of motion for the internal velocity field is obtained from the spatial components of the general relativistic conservation equation of energy and momentum,
\begin{equation}
	\partial_{\mu}M^{\mu i}+\partial_{\mu}P^{\mu i}=-\Gamma_{00}^i T^{00}, \label{eq:spatial_EMC}
\end{equation}
where, again, only $T^{00}$ terms were retained on the right-hand side due to (iii). 

As before, we start by evaluating the quantities on the left-hand side:
\begin{eqnarray}
	\partial_\mu M^{\mu i}&=&\partial_0(\rho\gamma^2cv^i)+\partial_j(\rho\gamma^2 v^iv^j)\nonumber\\
	&\simeq&\rho\frac{\partial v^i}{\partial t}+v^i\left[\frac{\partial\rho}{\partial t}+\partial_j(\rho v^j)\right]+\rho v^j\partial_j v^i\nonumber \\
	&=&\rho\left[\frac{\partial v^i}{\partial t}+(\bmath{v\cdot\nabla})v^i\right], \label{eq:M2}
\end{eqnarray}
where (iii) and Equation (\ref{eq:continuity_ap}) were employed. By making use of (i), (iii), (iv), and Equations (\ref{eq:FNC1})--(\ref{eq:FNC3}), one easily finds for the pressure tensor
\begin{eqnarray}
	\partial_\mu P^{\mu i}&=&\partial_0 p\left(\gamma^2\frac{v^i}{c}+g^{0i}\right)+\partial_k p\left(\gamma^2\frac{v^kv^i}{c^2}+g^{ki}\right)\nonumber\\
	&&\mskip10mu +p\left[\partial_0\left(\gamma^2\frac{v^i}{c}\right)+\partial_k\left(\gamma^2\frac{v^kv^i}{c^2}\right)+\partial_\mu g^{\mu i}\right]\nonumber\\
	&\simeq& \eta^{ki}\partial_kp+p\partial_\mu g^{\mu i}\nonumber\\
	&\simeq& \partial^i p, \label{eq:P2}
\end{eqnarray}
where in the second step we solely used (iii) and (iv). For the third step, the following expressions are needed: $R^{j\phantom{k}i}_{\phantom{j}k\phantom{i}j}=0$ and $R^{j\phantom{j}i}_{\phantom{j}j\phantom{i}k}\simeq0$. The former can be read off Equation (\ref{eq:Riemann}) and the latter is obtained from Equation (\ref{eq:Riemann}) by applying (i). Hence, it is easily seen that
\begin{eqnarray}
	\partial_\mu g^{\mu i}&=&\partial_0 g^{0i}+\partial_j g^{ji}\nonumber\\
	&\simeq&-\frac{4}{3}R^{0\phantom{k}i}_{\phantom{0}k\phantom{i}l}\partial_0(x^kx^l)-\frac{1}{3}\left(R^{j\phantom{k}i}_{\phantom{j}k\phantom{i}j}+R^{j\phantom{j}i}_{\phantom{j}j\phantom{i}k}\right)x^k\nonumber\\
	&\simeq& 0,
\end{eqnarray}
and thus Equation (\ref{eq:P2}) follows.

For the right-hand side of Equation (\ref{eq:spatial_EMC}) one needs to evaluate
\begin{eqnarray}
	\Gamma_{00}^{i}&=&g^{i\sigma}\left(\partial_0 g_{0\sigma}-\frac{1}{2}\partial_\sigma g_{00}\right)=-\frac{1}{2}g^{ij}\partial_jg_{00}\nonumber\\
	&\simeq& \frac{1}{2}\eta^{ij}R_{0k0l}\partial_j(x^kx^l)\nonumber\\
	&\simeq& -\frac{1}{2c^2}\ddot{h}^i_{\phantom{i}j}x^j, \label{eq:Gammai}
\end{eqnarray}
where in the second step we made use of Equation (\ref{eq:d0g00}) and 
\begin{equation}
\partial_0 g_{0i}\simeq-\frac{4}{3}R_{0kil}\partial_0(x^kx^l)=0.
\end{equation}
In the fourth step of Equation (\ref{eq:Gammai}), $R_{0k0l}=-\frac{1}{2c^2}\ddot{h}_{kl}$ has to be inserted, which is readily obtained from Equation (\ref{eq:Riemann}) by assuming (ii).

Thus, by inserting expressions (\ref{eq:M2}), (\ref{eq:P2}), and (\ref{eq:Gammai}) into Equation (\ref{eq:spatial_EMC}) we finally arrive at
\begin{equation}
	\rho\left[\frac{\partial\bmath{v}}{\partial t}+(\bmath{v\cdot\nabla})\bmath{v}\right]=-\nabla p+\bmath{f}_{\rm{GW}}, \label{eq:Euler_ap1}
\end{equation}
where the additional force per unit volume $\bmath{f}_{\rm{GW}}$ is defined by the components
\begin{equation}
	f_{\rm{GW}}^i=\frac{1}{2}\rho \ddot{h}^i_{\phantom{i}j} x^j.
\end{equation}
Employing Equation (\ref{eq:continuity_ap}) and using tensor product notation, Equation (\ref{eq:Euler_ap1}) is written as
\begin{equation}
\frac{\partial\rho\bmath{v}}{\partial t}+\bmath{\nabla\cdot}(\rho\bmath{v}\otimes\bmath{v})=-\nabla p+\bmath{f}_{\rm{GW}},
\end{equation}
which is the equation of motion to be derived in this subsection.

\section{Some properties of the gravitational wave tensor in transverse-traceless gauge}
\label{sec:app_2}

In order to clearly state the relevant definitions and equations employed in Sections \ref{sec:GW_radiation1} and \ref{sec:GW_radiation2}, we dedicate this appendix to a very brief review of some of the basic properties of the GW tensor $h_{\mu\nu}$ in TT gauge. For a textbook presentation of the subject see, e.g., \citet{Maggiore2008}.

In TT gauge (cf. Equation (\ref{eq:tt_gauge})), the GW tensor satisfies the linearized field equations in vacuo of the form
\begin{equation}
	\sq h_{ij}=0. \label{eq:field_eqs}
\end{equation}
Since $h_{ij}$ is traceless and symmetric (spin-2 operator), it can be expanded as
\begin{equation}
	h_{ij}=\sum_{m=-2}^{2}h_m\mathcal{Y}_{ij}^{2m}, \label{eq:expan_sph_harm}
\end{equation}
where the expansion coefficients $h_m$ are called the five independent spherical components of $h_{ij}$ and where the expansion was made in terms of the following basis for the five-dimensional space of traceless symmetric tensors:
\begin{eqnarray}
	\mathcal{Y}_{ij}^{2,\pm 2}&=&\sqrt{\frac{15}{32\pi}}\left( \begin{array}{ccc} 
	1 & \pm i & 0 \\ 
	\pm i & -1 & 0 \\
	0 & 0 & 0
	\end{array} \right)_{ij}, \label{eq:Y1}\\
	\mathcal{Y}_{ij}^{2,\pm 1}&=&\mp\sqrt{\frac{15}{32\pi}}\left( \begin{array}{ccc} 
	0 & 0 & 1 \\ 
	0 & 0 & \pm i \\
	1 & \pm i & 0
	\end{array} \right)_{ij},\\
	\mathcal{Y}_{ij}^{20}&=&\sqrt{\frac{5}{16\pi}}\left( \begin{array}{ccc} 
	-1 & 0 & 0 \\ 
	0 & -1 & 0 \\
	0 & 0 & 2
	\end{array} \right)_{ij}.\label{eq:Y3}
\end{eqnarray}
These tensors are related to spherical harmonics by the following relation:
\begin{equation}
	Y_{2m}(\theta,\phi)=\mathcal{Y}_{ij}^{2m}n_in_j,
\end{equation}
where $n_i$ denote the components of the radial unit vector
\begin{equation}
	\bmath{n}\equiv\bmath{e}_r=(\sin\theta\cos\phi,\sin\theta\sin\phi,\cos\theta). \label{eq:n}
\end{equation}
Furthermore, we mention the orthogonality relation that applies to these tensors:
\begin{equation}
	\sum_{ij}\mathcal{Y}_{ij}^{2m}\left(\mathcal{Y}_{ij}^{2m'}\right)^*=\frac{15}{8\pi}\delta^{mm'}.
\end{equation}
With the help of this orthogonality relation, Equation (\ref{eq:expan_sph_harm}) can be inverted to give
\begin{equation}
	h_m=\frac{8\pi}{15}h_{ij}\left(\mathcal{Y}_{ij}^{2m}\right)^*. \label{eq:inv_expan_sph_harm}
\end{equation}

Finally, we discuss the plane wave expansion of the GW tensor in TT gauge. Due to Equation (\ref{eq:field_eqs}), $h_{ij}$ can be expanded as
\begin{equation}
	h_{ij}(x)=\frac{1}{(2\pi)^3}\int d^3k\,\left[\mathcal{A}_{ij}(\bmath{k})e^{ikx}+\mathcal{A}_{ij}^*(\bmath{k})e^{-ikx}\right], \label{eq:pwe1}
\end{equation}
where $x\equiv(ct,\bmath{x})$ is used to denote spacetime points and where $k^\mu\equiv(\omega/c,\bmath{k})$ is the usual four-wave vector. The integral in Equation (\ref{eq:pwe1}) can alternatively be expressed as an integral over frequency, $\nu=\omega/2\pi$, and solid angle, $\Omega$,
\begin{equation}
	h_{ij}(x)=\frac{1}{c^3}\int_{0}^{\infty}d\nu\,\nu^2\int d\Omega\left[\mathcal{A}_{ij}(\nu,\bmath{n})e^{-i2\pi\nu(t-\bmath{n\cdot x}/c)}+c.c.\right], \label{eq:pwe2}
\end{equation}
since $d^3k=|\bmath{k}|^2 d|\bmath{k}| d\Omega=(2\pi/c)^3\nu^2 d\nu d\Omega$. Here $\bmath{n}$ was used to denote the propagation direction of the individual plane waves, $\bmath{n}=\bmath{k}/|\bmath{k}|$. We now define the polarization tensors
\begin{eqnarray}
	e_{ij}^+(\bmath{n})&=&\bmath{u}_i\bmath{u}_j-\bmath{v}_i\bmath{v}_j, \label{eq:e1}\\
	e_{ij}^\times(\bmath{n})&=&\bmath{u}_i\bmath{v}_j+\bmath{u}_j\bmath{v}_i, \label{eq:e2}
\end{eqnarray}
where $\bmath{n}$ given by Equation (\ref{eq:n}), $\bmath{u}$, and $\bmath{v}$ are pairwise orthonormal vectors,
\begin{eqnarray}
	\bmath{u}\equiv\bmath{e}_{\Theta}&=&(\cos\Theta\cos\phi,\cos\Theta\sin\phi,-\sin\Theta)\label{eq:u}\\
	\bmath{v}\equiv\bmath{e}_{\phi}&=&(-\sin\phi,\cos\phi,0).\label{eq:v}
\end{eqnarray}
From this definition, it is clear that the polarization tensors represent transversal polarization states and that these polarization states are independent. The latter point is also reflected by the fact that
\begin{equation}
	e_{ij}^{A}(\bmath{n})e^{A',ij}(\bmath{n})=2\delta^{A,A'},
\end{equation}
which can be proved explicitly. Here $A=\{+,\times\}$ labels the polarization states. Since applying the gauge conditions (\ref{eq:tt_gauge}) to Equation (\ref{eq:pwe1}) yields $k^i\mathcal{A}_{ij}(\bmath{k})=0$, the tensor components $\mathcal{A}_{ij}$ can be expanded in terms of the polarization tensors,
\begin{equation}
	\mathcal{A}_{ij}(\nu,\bmath{n})=\frac{c^3}{\nu^2}\sum_{A}h_A(\nu,\bmath{n})e_{ij}^A(\bmath{n}), \label{eq:def_fourier_ampl}
\end{equation}
where the expansion coefficients $h_A(\nu,\bmath{n})$ are called the Fourier amplitudes of $h_{ij}$. Substituting Equation (\ref{eq:def_fourier_ampl}) into Equation (\ref{eq:pwe2}) and extending the frequency integration domain to minus infinity by setting $h_A(-\nu,\bmath{n})\equiv h_A^*(\nu,\bmath{n})$ finally yields
\begin{equation}
	h_{ij}(t,\bmath{x})=\sum_{A}\int_{-\infty}^{\infty} d\nu\int d\Omega\,h_A(\nu,\bmath{n})e_{ij}^A(\bmath{n})e^{-i2\pi\nu(t-\bmath{n\cdot x}/c)}. \label{eq:pwe3}
\end{equation}
This is the desired form of the plane wave expansion for the GW tensor employed in Section \ref{sec:GW_radiation2}.




\bibliography{Library.bib}{}

\begin{thebibliography}{77}
\expandafter\ifx\csname natexlab\endcsname\relax\def\natexlab#1{#1}\fi

\bibitem[{{Abbott} {et~al.}(2009){Abbott}, {Abbott}, {Acernese}, {Adhikari},
  {Ajith}, {Allen}, {Allen}, {Alshourbagy}, {Amin}, {Anderson}, \&
  et~al.}]{Abbott2009}
{Abbott}, B.~P., {et~al.} 2009, \nat, 460, 990

\bibitem[{{Aizenman} \& {Smeyers}(1977)}]{Aizenman1977}
{Aizenman}, M.~L., \& {Smeyers}, P. 1977, \apss, 48, 123

\bibitem[{{Allen}(1988)}]{Allen1988}
{Allen}, B. 1988, \prd, 37, 2078

\bibitem[{{Allen}(1997)}]{Allen1997}
{Allen}, B. 1997, in Some Topics on General Relativity and Gravitational
  Radiation, ed. J.~A. {Miralles}, J.~A. {Morales}, \& D.~{Saez}, Proc. of the
  ``Spanish Relativity Meeting '96'', Valencia, Spain (Paris: Editions
  Frontieres), 3

\bibitem[{{Allen} \& {Romano}(1999)}]{Allen1999}
{Allen}, B., \& {Romano}, J.~D. 1999, \prd, 59, 102001

\bibitem[{{Appourchaux} {et~al.}(2010){Appourchaux}, {Belkacem}, {Broomhall},
  {Chaplin}, {Gough}, {Houdek}, {Provost}, {Baudin}, {Boumier}, {Elsworth},
  {Garc{\'{\i}}a}, {Andersen}, {Finsterle}, {Fr{\"o}hlich}, {Gabriel}, {Grec},
  {Jim{\'e}nez}, {Kosovichev}, {Sekii}, {Toutain}, \&
  {Turck-Chi{\`e}ze}}]{Appourchaux2010}
{Appourchaux}, T., {et~al.} 2010, \aapr, 18, 197

\bibitem[{{Baglin} {et~al.}(2006){Baglin}, {Michel}, {Auvergne}, \& {The COROT
  Team}}]{Baglin2006}
{Baglin}, A., {Michel}, E., {Auvergne}, M., \& {The COROT Team}. 2006, in ESA
  Special Publication, Vol. 624, Proceedings of SOHO 18/GONG 2006/HELAS I,
  Beyond the spherical Sun, ed. K.~{Fletcher} \& M.~{Thompson} (Noordwijk:
  ESA), 34

\bibitem[{{Balmforth}(1992)}]{Balmforth1992}
{Balmforth}, N.~J. 1992, \mnras, 255, 639

\bibitem[{{Bar-Kana}(1994)}]{Bar-Kana1994}
{Bar-Kana}, R. 1994, \prd, 50, 1157

\bibitem[{{Belkacem} {et~al.}(2008){Belkacem}, {Samadi}, {Goupil}, \&
  {Dupret}}]{Belkacem2008}
{Belkacem}, K., {Samadi}, R., {Goupil}, M., \& {Dupret}, M. 2008, \aap, 478,
  163

\bibitem[{{Belkacem} {et~al.}(2009){Belkacem}, {Samadi}, {Goupil}, {Dupret},
  {Brun}, \& {Baudin}}]{Belkacem2009}
{Belkacem}, K., {Samadi}, R., {Goupil}, M.~J., {Dupret}, M.~A., {Brun}, A.~S.,
  \& {Baudin}, F. 2009, \aap, 494, 191

\bibitem[{{Belkacem} {et~al.}(2006{\natexlab{a}}){Belkacem}, {Samadi},
  {Goupil}, \& {Kupka}}]{Belkacem2006a}
{Belkacem}, K., {Samadi}, R., {Goupil}, M.~J., \& {Kupka}, F.
  2006{\natexlab{a}}, \aap, 460, 173

\bibitem[{{Belkacem} {et~al.}(2006{\natexlab{b}}){Belkacem}, {Samadi},
  {Goupil}, {Kupka}, \& {Baudin}}]{Belkacem2006b}
{Belkacem}, K., {Samadi}, R., {Goupil}, M.~J., {Kupka}, F., \& {Baudin}, F.
  2006{\natexlab{b}}, \aap, 460, 183

\bibitem[{{Boughn} \& {Kuhn}(1984)}]{Boughn1984}
{Boughn}, S.~P., \& {Kuhn}, J.~R. 1984, \apj, 286, 387

\bibitem[{{Brustein} {et~al.}(1995){Brustein}, {Gasperini}, {Giovannini}, \&
  {Veneziano}}]{Brustein1995}
{Brustein}, R., {Gasperini}, M., {Giovannini}, M., \& {Veneziano}, G. 1995,
  \plb, 361, 45

\bibitem[{{Buonanno} {et~al.}(1997){Buonanno}, {Maggiore}, \&
  {Ungarelli}}]{Buonanno1997}
{Buonanno}, A., {Maggiore}, M., \& {Ungarelli}, C. 1997, \prd, 55, 3330

\bibitem[{{Caldwell} \& {Allen}(1992)}]{Caldwell1992}
{Caldwell}, R.~R., \& {Allen}, B. 1992, \prd, 45, 3447

\bibitem[{{Caldwell} {et~al.}(1996){Caldwell}, {Battye}, \&
  {Shellard}}]{Caldwell1996}
{Caldwell}, R.~R., {Battye}, R.~A., \& {Shellard}, E.~P.~S. 1996, \prd, 54,
  7146

\bibitem[{{Christensen-Dalsgaard}(2008)}]{ChristensenDalsgaard2008}
{Christensen-Dalsgaard}, J. 2008, \apss, 316, 113

\bibitem[{{Christensen-Dalsgaard} {et~al.}(2009){Christensen-Dalsgaard},
  {Arentoft}, {Brown}, {Gilliland}, {Kjeldsen}, {Borucki}, \&
  {Koch}}]{Christensen-Dalsgaard2009}
{Christensen-Dalsgaard}, J., {Arentoft}, T., {Brown}, T.~M., {Gilliland},
  R.~L., {Kjeldsen}, H., {Borucki}, W.~J., \& {Koch}, D. 2009, Commun.
  Asteroseismol., 158, 328

\bibitem[{{Christensen-Dalsgaard} {et~al.}(1996){Christensen-Dalsgaard},
  {Dappen}, {Ajukov}, {Anderson}, {Antia}, {Basu}, {Baturin}, {Berthomieu},
  {Chaboyer}, {Chitre}, {Cox}, {Demarque}, {Donatowicz}, {Dziembowski},
  {Gabriel}, {Gough}, {Guenther}, {Guzik}, {Harvey}, {Hill}, {Houdek},
  {Iglesias}, {Kosovichev}, {Leibacher}, {Morel}, {Proffitt}, {Provost},
  {Reiter}, {Rhodes}, {Rogers}, {Roxburgh}, {Thompson}, \&
  {Ulrich}}]{ChristensenDalsgaard1996}
{Christensen-Dalsgaard}, J., {et~al.} 1996, Science, 272, 1286

\bibitem[{{Copeland} {et~al.}(2004){Copeland}, {Myers}, \&
  {Polchinski}}]{Copeland2004}
{Copeland}, E.~J., {Myers}, R.~C., \& {Polchinski}, J. 2004, \jhep, 06, 013

\bibitem[{{Cyburt} {et~al.}(2005){Cyburt}, {Fields}, {Olive}, \&
  {Skillman}}]{Cyburt2005}
{Cyburt}, R.~H., {Fields}, B.~D., {Olive}, K.~A., \& {Skillman}, E. 2005, \aph,
  23, 313

\bibitem[{{Damour} \& {Vilenkin}(2001)}]{Damour2001}
{Damour}, T., \& {Vilenkin}, A. 2001, \prd, 64, 064008

\bibitem[{{Damour} \& {Vilenkin}(2005)}]{Damour2005}
---. 2005, \prd, 71, 063510

\bibitem[{{Depies} \& {Hogan}(2007)}]{DePies2007}
{Depies}, M.~R., \& {Hogan}, C.~J. 2007, \prd, 75, 125006

\bibitem[{{Dupret}(2002)}]{Dupret2002}
{Dupret}, M.~A. 2002, Bull. Soc. R. Sci. Li{\`e}ge, 71, 249

\bibitem[{{Dvali} \& {Vilenkin}(2004)}]{Dvali2004}
{Dvali}, G., \& {Vilenkin}, A. 2004, \jcap, 03, 010

\bibitem[{{Dyson}(1969)}]{Dyson1969}
{Dyson}, F.~J. 1969, \apj, 156, 529

\bibitem[{{Enoki} {et~al.}(2004){Enoki}, {Inoue}, {Nagashima}, \&
  {Sugiyama}}]{Enoki2004}
{Enoki}, M., {Inoue}, K.~T., {Nagashima}, M., \& {Sugiyama}, N. 2004, \apj,
  615, 19

\bibitem[{{Farmer} \& {Phinney}(2003)}]{Farmer2003}
{Farmer}, A.~J., \& {Phinney}, E.~S. 2003, \mnras, 346, 1197

\bibitem[{{Forward} {et~al.}(1961){Forward}, {Zipoy}, \& {Weber}}]{Forward1961}
{Forward}, R.~L., {Zipoy}, D., \& {Weber}, J. 1961, \nat, 189, 473

\bibitem[{{Goldreich} \& {Keeley}(1977)}]{Goldreich1977}
{Goldreich}, P., \& {Keeley}, D.~A. 1977, \apj, 212, 243

\bibitem[{{Goldreich} {et~al.}(1994){Goldreich}, {Murray}, \&
  {Kumar}}]{Goldreich1994}
{Goldreich}, P., {Murray}, N., \& {Kumar}, P. 1994, \apj, 424, 466

\bibitem[{{Gough}(1985)}]{Gough1985}
{Gough}, D.~O. 1985, in ESA Special Publication, Vol. 235, Future Missions in
  Solar, Heliospheric \& Space Plasma Physics, ed. E.~{Rolfe} \& B.~{Battrick}
  (Noordwijk: ESA Scientific \& Technical Publications Branch), 183

\bibitem[{{Grigahc{\`e}ne} {et~al.}(2005){Grigahc{\`e}ne}, {Dupret}, {Gabriel},
  {Garrido}, \& {Scuflaire}}]{Grigahcene2005}
{Grigahc{\`e}ne}, A., {Dupret}, M., {Gabriel}, M., {Garrido}, R., \&
  {Scuflaire}, R. 2005, \aap, 434, 1055

\bibitem[{{Jackson} {et~al.}(2005){Jackson}, {Jones}, \&
  {Polchinski}}]{Jackson2005}
{Jackson}, M.~G., {Jones}, N.~T., \& {Polchinski}, J. 2005, \jhep, 10, 013

\bibitem[{{Jenet} {et~al.}(2006){Jenet}, {Hobbs}, {van Straten}, {Manchester},
  {Bailes}, {Verbiest}, {Edwards}, {Hotan}, {Sarkissian}, \& {Ord}}]{Jenet2006}
{Jenet}, F.~A., {et~al.} 2006, \apj, 653, 1571

\bibitem[{{Jeong} \& {Smoot}(2005)}]{Jeong2005}
{Jeong}, E., \& {Smoot}, G.~F. 2005, \apj, 624, 21

\bibitem[{{Jones} {et~al.}(2002){Jones}, {Stoica}, \& {Tye}}]{Jones2002}
{Jones}, N., {Stoica}, H., \& {Tye}, S. 2002, \jhep, 07, 051

\bibitem[{{Jones} {et~al.}(2003){Jones}, {Stoica}, \& {Tye}}]{Jones2003}
{Jones}, N.~T., {Stoica}, H., \& {Tye}, S. 2003, \plb, 563, 6

\bibitem[{{Khosroshahi} \& {Sobouti}(1997)}]{Khosroshahi1997}
{Khosroshahi}, H.~G., \& {Sobouti}, Y. 1997, \aap, 321, 1024

\bibitem[{{Kibble}(1976)}]{Kibble1976}
{Kibble}, T.~W.~B. 1976, \jpa, 9, 1387

\bibitem[{{Kumar} {et~al.}(1996){Kumar}, {Quataert}, \& {Bahcall}}]{Kumar1996}
{Kumar}, P., {Quataert}, E.~J., \& {Bahcall}, J.~N. 1996, \apjl, 458, L83

\bibitem[{{Kuroyanagi} {et~al.}(2009){Kuroyanagi}, {Chiba}, \&
  {Sugiyama}}]{Kuroyanagi2009}
{Kuroyanagi}, S., {Chiba}, T., \& {Sugiyama}, N. 2009, \prd, 79, 103501

\bibitem[{{Maggiore}(2000)}]{Maggiore2000}
{Maggiore}, M. 2000, \physrep, 331, 283

\bibitem[{{Maggiore}(2008)}]{Maggiore2008}
---. 2008, Gravitational Waves Vol.1: Theory and Experiments (New York, NY:
  Oxford Univ. Press)

\bibitem[{{Manasse} \& {Misner}(1963)}]{ManasseMisner1963}
{Manasse}, F.~K., \& {Misner}, C.~W. 1963, \jmp, 4, 735

\bibitem[{{Mandic} \& {Buonanno}(2006)}]{Mandic2006}
{Mandic}, V., \& {Buonanno}, A. 2006, \prd, 73, 063008

\bibitem[{{Misner} {et~al.}(1973){Misner}, {Thorne}, \&
  {Wheeler}}]{MisnerThorneWheeler1973}
{Misner}, C.~W., {Thorne}, K.~S., \& {Wheeler}, J.~A. 1973, {Gravitation} (San
  Francisco, CA: Freeman)

\bibitem[{{Nelemans}(2009)}]{Nelemans2009}
{Nelemans}, G. 2009, \cqg, 26, 094030

\bibitem[{{Nelemans} {et~al.}(2001){Nelemans}, {Yungelson}, \& {Portegies
  Zwart}}]{Nelemans2001}
{Nelemans}, G., {Yungelson}, L.~R., \& {Portegies Zwart}, S.~F. 2001, \aap,
  375, 890

\bibitem[{{Nelemans} {et~al.}(2004){Nelemans}, {Yungelson}, \& {Portegies
  Zwart}}]{Nelemans2004}
---. 2004, \mnras, 349, 181

\bibitem[{{Nelemans} {et~al.}(2005){Nelemans}, {Napiwotzki}, {Karl}, {Marsh},
  {Voss}, {Roelofs}, {Izzard}, {Montgomery}, {Reerink}, {Christlieb}, \&
  {Reimers}}]{Nelemans2005}
{Nelemans}, G., {et~al.} 2005, \aap, 440, 1087

\bibitem[{{{\"O}lmez} {et~al.}(2010){{\"O}lmez}, {Mandic}, \&
  {Siemens}}]{Oelmez2010}
{{\"O}lmez}, S., {Mandic}, V., \& {Siemens}, X. 2010, \prd, 81, 104028

\bibitem[{{Pogosian} {et~al.}(2004){Pogosian}, {Wyman}, \&
  {Wasserman}}]{Pogosian2004}
{Pogosian}, L., {Wyman}, M., \& {Wasserman}, I. 2004, \jcap, 09, 008

\bibitem[{{Roelofs} {et~al.}(2007{\natexlab{a}}){Roelofs}, {Groot}, {Benedict},
  {McArthur}, {Steeghs}, {Morales-Rueda}, {Marsh}, \&
  {Nelemans}}]{Roelofs2007a}
{Roelofs}, G.~H.~A., {Groot}, P.~J., {Benedict}, G.~F., {McArthur}, B.~E.,
  {Steeghs}, D., {Morales-Rueda}, L., {Marsh}, T.~R., \& {Nelemans}, G.
  2007{\natexlab{a}}, \apj, 666, 1174

\bibitem[{{Roelofs} {et~al.}(2006){Roelofs}, {Groot}, {Nelemans}, {Marsh}, \&
  {Steeghs}}]{Roelofs2006}
{Roelofs}, G.~H.~A., {Groot}, P.~J., {Nelemans}, G., {Marsh}, T.~R., \&
  {Steeghs}, D. 2006, \mnras, 371, 1231

\bibitem[{{Roelofs} {et~al.}(2007{\natexlab{b}}){Roelofs}, {Nelemans}, \&
  {Groot}}]{Roelofs2007b}
{Roelofs}, G.~H.~A., {Nelemans}, G., \& {Groot}, P.~J. 2007{\natexlab{b}},
  \mnras, 382, 685

\bibitem[{{Ruiter} {et~al.}(2010){Ruiter}, {Belczynski}, {Benacquista},
  {Larson}, \& {Williams}}]{Ruiter2010}
{Ruiter}, A.~J., {Belczynski}, K., {Benacquista}, M., {Larson}, S.~L., \&
  {Williams}, G. 2010, \apj, 717, 1006

\bibitem[{{Sahni}(1990)}]{Sahni1990}
{Sahni}, V. 1990, \prd, 42, 453

\bibitem[{{Samadi} \& {Goupil}(2001)}]{Samadi2001a}
{Samadi}, R., \& {Goupil}, M. 2001, \aap, 370, 136

\bibitem[{{Samadi} {et~al.}(2003{\natexlab{a}}){Samadi}, {Nordlund}, {Stein},
  {Goupil}, \& {Roxburgh}}]{Samadi2003b}
{Samadi}, R., {Nordlund}, {\AA}., {Stein}, R.~F., {Goupil}, M.~J., \&
  {Roxburgh}, I. 2003{\natexlab{a}}, \aap, 404, 1129

\bibitem[{{Samadi} {et~al.}(2003{\natexlab{b}}){Samadi}, {Nordlund}, {Stein},
  {Goupil}, \& {Roxburgh}}]{Samadi2003a}
---. 2003{\natexlab{b}}, \aap, 403, 303

\bibitem[{{Sarangi} \& {Tye}(2002)}]{Sarangi2002}
{Sarangi}, S., \& {Tye}, S. 2002, \plb, 536, 185

\bibitem[{{Sesana} {et~al.}(2004){Sesana}, {Haardt}, {Madau}, \&
  {Volonteri}}]{Sesana2004}
{Sesana}, A., {Haardt}, F., {Madau}, P., \& {Volonteri}, M. 2004, \apj, 611,
  623

\bibitem[{{Sesana} {et~al.}(2005){Sesana}, {Haardt}, {Madau}, \&
  {Volonteri}}]{Sesana2005}
---. 2005, \apj, 623, 23

\bibitem[{{Siegel} \& {Roth}(2010)}]{Siegel2010}
{Siegel}, D.~M., \& {Roth}, M. 2010, \mnras, 408, 1742

\bibitem[{{Siemens} {et~al.}(2007){Siemens}, {Mandic}, \&
  {Creighton}}]{Siemens2007}
{Siemens}, X., {Mandic}, V., \& {Creighton}, J. 2007, \prl, 98, 111101

\bibitem[{{Smith} {et~al.}(2006){Smith}, {Pierpaoli}, \&
  {Kamionkowski}}]{Smith2006}
{Smith}, T.~L., {Pierpaoli}, E., \& {Kamionkowski}, M. 2006, \prl, 97, 021301

\bibitem[{{Starobinski{\v \i}}(1979)}]{Starobinskii1979}
{Starobinski{\v \i}}, A.~A. 1979, \sjetpl, 30, 719

\bibitem[{{Timpano} {et~al.}(2006){Timpano}, {Rubbo}, \&
  {Cornish}}]{Timpano2006}
{Timpano}, S.~E., {Rubbo}, L.~J., \& {Cornish}, N.~J. 2006, \prd, 73, 122001

\bibitem[{{Unno} {et~al.}(1989){Unno}, {Osaki}, {Ando}, {Saio}, \&
  {Shibahashi}}]{Unno}
{Unno}, W., {Osaki}, Y., {Ando}, H., {Saio}, H., \& {Shibahashi}, H. 1989,
  {Nonradial oscillations of stars}, 2nd edn. (Tokyo: Univ. of Tokyo Press)

\bibitem[{{Vilenkin}(1981)}]{Vilenkin1981}
{Vilenkin}, A. 1981, \plb, 107, 47

\bibitem[{{Weber}(1967)}]{Weber1967}
{Weber}, J. 1967, \prl, 18, 498

\bibitem[{{Willems} {et~al.}(2007){Willems}, {Kalogera}, {Vecchio}, {Ivanova},
  {Rasio}, {Fregeau}, \& {Belczynski}}]{Willems2007}
{Willems}, B., {Kalogera}, V., {Vecchio}, A., {Ivanova}, N., {Rasio}, F.~A.,
  {Fregeau}, J.~M., \& {Belczynski}, K. 2007, \apj, 665, L59

\bibitem[{{Zimmerman} \& {Hellings}(1980)}]{Zimmerman1980}
{Zimmerman}, R.~L., \& {Hellings}, R.~W. 1980, \apj, 241, 475

\end{thebibliography}
\bibliographystyle{apj.bst}

\clearpage

\end{document}